\documentclass{optica-article}

\journal{opticajournal} % for journals or Optica Open

\articletype{Research Article}

\usepackage{lineno}
\usepackage{color}
\usepackage{xcolor}
\usepackage{CJK}
\modulolinenumbers[5]

\begin{document}
\begin{CJK*}{UTF8}{gbsn}
\title{Design and optimization of multilevel zone plates for flying focus applications}

\author{
Zhengkun Li,\authormark{1}
Yisong Zhou,\authormark{1}
Changbo Fu,\authormark{1,*}
Yugang Ma,\authormark{1}
}

\address{\authormark{1}Key Lab of Nuclear Physics and Ion-beam Application (MoE), Institute of Modern Physics, Fudan University, Shanghai 200433,  China\\
% \email{25110200015@m.fudan.edu.cn}
\email{\authormark{*}cbfu@fudan.edu.cn}
}

 %% email address is required; see note below about the corresponding author designation

% use {asbstract*} to suppress the copyright line. Copyright information will be added in production

\begin{abstract*} 
Flying focus pulses, characterized by programmable velocities decoupled from the group velocity, offer remarkable potential for high-intensity laser-matter interactions. However, achieving low-cost manufacturing while simultaneously suppressing higher-order foci and maintaining high efficiency remains a key challenge. Here, we propose a modulated width zone plate with $2^{M}$ steps (MWZP-$2^{M}$S). By employing Monte Carlo sampling of sinusoidal random widths, the MWZP-8S design statistically mimics a Gabor zone plate, suppressing higher-order foci by four orders of magnitude. Furthermore, the multilevel phase modulation converts opaque regions into constructive phase-shifting zones, significantly boosting the focusing efficiency to 73.7$\%$ when $M=3$. Numerical simulations demonstrate that the focal velocity is flexibly tunable across subluminal and superluminal regimes while remaining stable over 100 Rayleigh lengths. This scalable and cost-effective scheme provides a robust platform for flying focus applications in EUV generation and X-ray optics.

\end{abstract*}

%%%%%%%%%%%%%%%%%%%%%%%%%%  body  %%%%%%%%%%%%%%%%%%%%%%%%%%
\section{Introduction}

Spatiotemporal structured light, whose spatial and temporal structures can be tuned simultaneously \cite{Akturk_2010,Zhu:05,PhysRevA.99.023856}, 
serves as a powerful enabler of advanced optical technologies, from the radiation of terahertz waves \cite{toth_tilted_2023} and the creation of laser beams with time-varying orbital angular momentum (OAM) \cite{doi:10.1126/science.aaw9486} to the generation of flying-focus pulses\cite{Sainte-Marie2017,Froula2018,Pigeon2024}.
% including attosecond pulse sequences\cite{PhysRevLett.108.113904,wheeler_attosecond_2012}, terahertz wave generation\cite{toth_tilted_2023}, and the creation of flying focus pulses\cite{Sainte-Marie2017,Froula2018,Pigeon2024}. 
% % \cite{Froula2018,Jolly:20}
The flying focus approach, as a particular spatiotemporal light scheme, has been employed to optimize diverse laser-based applications necessitating velocity matching or extended interaction lengths\cite{Trends2025}, including laser-wakefield acceleration\cite{PhysRevLett.124.134802,Caizergues2020,PhysRevX.9.031044}, pump-probe synchronization in soft X-ray lasers\cite{SoftXray_laser_2023}, and the creation of long, contiguous plasma channels \cite{PhysRevA.97.033835,PhysRevLett.120.225001,Franke:19}.
% nonlinear Thomson scattering \cite{PhysRevE.105.065201}, and Compton scattering \cite{PhysRevA.103.012215}.

% However, advancing these applications further requires schemes that address two practical challenges: compatibility with standard nanofabrication processes for scalable and cost-effective manufacturing, and the ability to focus light into a single dominant spot while eliminating disruptive higher-order foci.
% However, the fulfillment of this considerable potential requires the development of versatile and efficient optical configurations for flying focus generation.

 There have been many methods to realize flying focus\cite{Froula2018,Spatial_and_temporal_2019,PhysRevLett.120.225001,Jolly:20,PhysRevLett.124.134802,Pigeon2024}. 
Pioneering flying focus generation schemes exploit chromatic diffractive lenses with radially gradient groove density to focus frequency-chirped laser pulses\cite{Froula2018,Spatial_and_temporal_2019,PhysRevLett.120.225001}. 
Nevertheless, the manufacturing of such diffractive optical elements demands high-precision and costly micromachining technologies, including single-point diamond turning and electron-beam lithography\cite{Fleming:97}.
Ambat et al. \cite{Ambat:23} and Markland et al. \cite{Markland:26} demonstrated programmable flying-focus trajectories using an axiparabola combined with a deformable mirror (DM) and a spatial light modulator (SLM), while Blum et al. \cite{Blum:26} employed a 52-actuator DM to program spherical aberration and focal elongation on shot. The DM-based schemes \cite{Markland:26,Blum:26} are constrained by the DM sag, which limits the achievable radial group delay. In addition, the spherical-aberration-based focusing of the axiparabola produces on-axis intensity modulation and a longitudinally varying focal spot \cite{Ambat:23,Markland:26}. 
 % The pioneering approach uses a chromatic diffractive lens with a radially varying groove density to focus a frequency-chirped laser pulse\cite{Froula2018,Spatial_and_temporal_2019,PhysRevLett.120.225001}. The fabrication of such diffractive lenses typically requires high-precision and costly micromachining techniques, such as single-point diamond turning or electron-beam lithography\cite{Fleming:97}.

% Ambat et al. \cite{Ambat:23} and Markland et al. \cite{Markland:26} demonstrate programmable flying focus trajectories using an axiparabola combined with a deformable mirror and spatial light modulator. However, this approach requires
% multiple optical components which are complex and costly.
% To enable on-shot programmable control of spherical aberration and focal elongation, another scheme\cite{Blum:26} leverages a complex deformable mirror with even 52 actuators. This functionality, moreover, mandates that the mirror be placed at a considerable distance upstream from the final focusing optic, resulting in a bulky and integration-sensitive optical layout.
% {\lzk Another work utilizes a specialized stretched off-axis paraboloid (sOAP) in combination with a diffraction grating to generate structured flying focus trains \cite{ComptonZhao2025}. Although this method enables customized flying focus profiles for extreme ultraviolet light generation, the specially tailored sOAP requires sophisticated fabrication techniques and suffers from high manufacturing costs.}

Here raises the practically important question: can flying focus be realized using diffractive optical element (DOE)\cite{:/ActPhotonSin/doi/10.3788/gzxb20265506.0622001}? DOEs can be fabricated with mature semiconductor photolithographic and etching processes\cite{Sankur2000}, which provide multilevel phase profiles with micrometer-to-wavelength-scale features\cite{ZHANG2018387}. This mature fabrication basis makes DOEs a practical route to a flying focus.

Advancing flying focus applications would benefit from
cost-effective manufacturing and Fresnel zone plates (FZP) could provide this benefit
but require the ability to focus light into a single dominant spot while eliminating
disruptive higher-order foci.  The modulated width zone plate (MWZP)\cite{ZHANG2018387}, with a radially modulated sinusoidal transmittance to simulate the Gabor zone plate (GZP)\cite{Gabor:48, Rogers:50}, exhibits low focusing efficiency by functioning as a binary amplitude mask. To overcome these difficulties, we combine the idea of multilevel zone plates \cite{DiFabrizio1999} with MWZP and design the zone plates named MWZP-2steps (MWZP-2S), MWZP-4steps (MWZP-4S), and MWZP-8steps (MWZP-8S) to enhance focusing efficiency and improve the suppression of higher-order foci.
 % Our designed multilevel zone plates suppresses higher-order foci and enhances focusing efficiency.
 % Fresnel zone plates' (FZP) high-order foci introduce significant background noise, making their suppression essential. 
 
 % The Gabor zone plate (GZP), with a radially continuous sinusoidal transmittance, theoretically eliminates all high-order foci\cite{Gabor:48, Rogers:50}, but its fabrication remains technologically difficult with nanofabrication methods.  
 % To overcome these limitations, we adopt the modulated width zone plate (MWZP)\cite{ZHANG2018387}, a binary amplitude mask with azimuthal symmetry, to suppress all high-order foci in chirped Gaussian pulse applications.

In this study, 
we propose a DOE-based flying focus scheme, which suppresses higher-order foci,
and improves focusing efficiency of flying focus.
% This novel approach introduces a sinusoidal modulation of the zone widths, structurally approximating the statistical transparency profile of the GZP. 
% Moreover, the focus performance improves significantly as the number of illuminated zones increases,  thereby enabling a well-defined single focus.
In the following sections,
we first introduce the concept of flying focus and its important characteristics. 
Next, the design of the zone plates and the modified Fresnel integrals are illustrated. Finally, numerical simulations of the DOE-based flying focus are presented, along with a discussion of fabrication stability.

%==============================================================
%==============================================================
%==============================================================
%==============================================================
\section{Methods}

\begin{figure}[htbp]
\centering\includegraphics[width=0.7\linewidth]{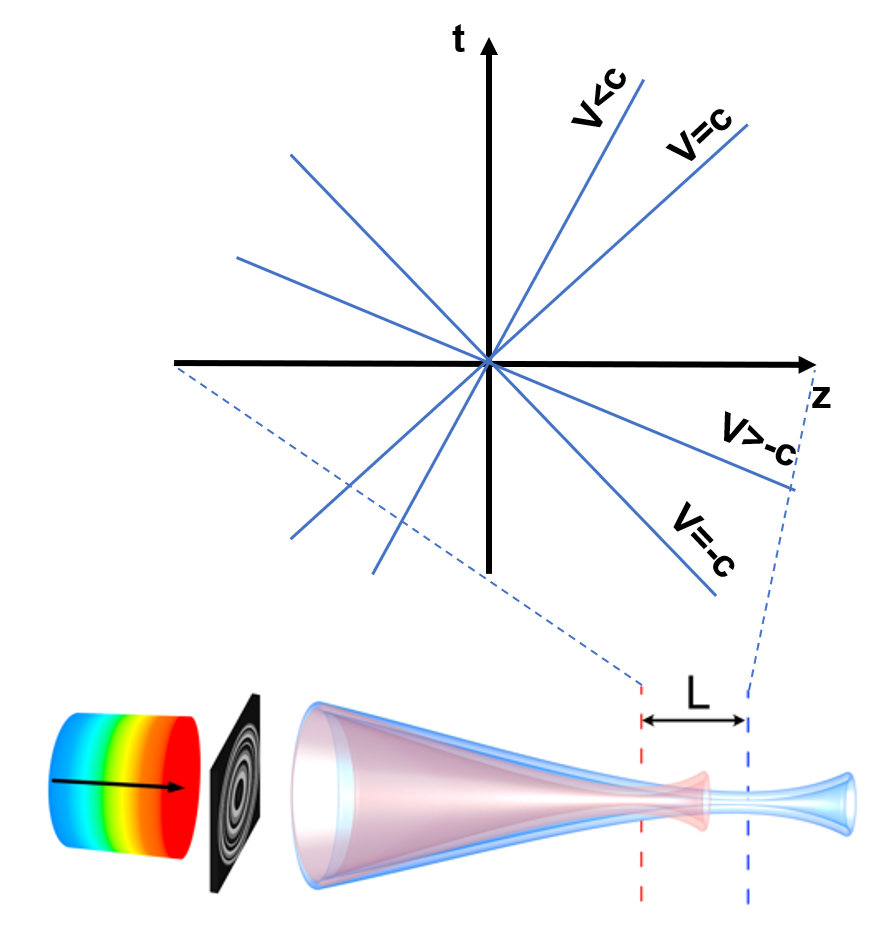}
\caption{ Schematic of the generation of flying focus pulses with tunable velocity using a designed zone plate. Bottom: A chirped laser pulse is focused by the zone plate, which exhibits longitudinal chromatic aberration, creating an extended focal region. Top: Corresponding space-time ($t-z$) trajectories. By tuning the pulse chirp (from positive to strongly negative), the peak intensity can propagate forward at a subluminal velocity ($v < c$), or backward at a superluminal velocity ($v > -c$).
}
\label{flying focus concept}
\end{figure}

\subsection{From physical demands to flying focus generation}

Many laser-based applications critically rely on the propagation of light over long distances. A primary requirement is the ability to maintain a focus over a longitudinal distance far exceeding the conventional Rayleigh length (e.g., on the order of \(10\) mm). Simultaneously, this extended region must preserve a nearly diffraction-limited minimum spot size (e.g., \(\sim 10\) \(\mu\)m), to maintain high intensity throughout the focus range. 

Beyond the spatial and intensity constraints, precise control over the propagation speed of the peak intensity is equally critical. In many physical processes, the velocity of the laser intensity must be decoupled from the natural group velocity of light. For instance, in longitudinally pumped seeded soft-X-ray lasers \cite{Kabacinski2023}, a forward-propagating focus at light speed c is required to synchronize the pump beam with X-ray pulses.
% the moving plasma gain region. 
Conversely, a photon accelerator \cite{PhysRevLett.61.337,PhysRevLett.123.124801} requires a backward-propagating focus at vacuum light speed to prevent the accelerated photons from outpacing the ionization front. 

Such stringent requirements-a long focal range with a small spot size, coupled with a programmable velocity-cannot be satisfied by conventional optics with no chirp. To address this challenge, the concept of the flying focus has been proposed. The flying focus utilizes a chromatic focusing system combined with a broadband laser pulse (for example, 800nm central wavelength with a bandwidth of approximately \(20\) nm and a chirped pulse duration about \(200\) fs) to propagate the focus over an extended distance with a self-similar focus profile, while decoupling the focal velocity from the group velocity of the light \cite{Spatial_and_temporal_2019}.

% Conventional optics with no chirp cannot decouple the focal depth from the spot size, but this limitation can be overcome by the concept of the flying focus \cite{Spatial_and_temporal_2019}. 
% The flying focus leverages a chromatic focusing system to map the spectral components of the broadband pulse to different longitudinal locations, allowing the focus to propagate over an extended distance with a self-similar profile. 

As illustrated in Fig. \ref{flying focus concept}, the core physical mechanism relies on longitudinal chromatic aberration, which maps different spectral components of a broadband pulse to different longitudinal locations. For a system with normal dispersion, the longer-wavelength (red) light focuses closer to the optic, whereas the shorter-wavelength (blue) light focuses further away. The longitudinal focal range, determined by the laser bandwidth (\(\Delta\lambda = \lambda_{\mathrm{r}} - \lambda_{\mathrm{b}}\)), is given by \(\Delta z \simeq f_0 \Delta\lambda / \lambda_0\), where \(f_0\) is the nominal focal length corresponding to the central wavelength \(\lambda_0\). Throughout this extended focal region, the minimum spot size remains approximately \(w_0 \simeq \lambda_0 f_0 / (\pi R)\), where \(R\) is the radius of the incident beam, meeting the spatial and spot size requirements. 

The most distinct feature of the flying focus is that its focal velocity is no longer constrained by the group velocity. This arbitrary velocity control is achieved by tuning the pulse chirp, which determines the temporal order in which the wavelengths reach their respective foci. As depicted in the space-time \((t-z)\) diagram in Fig. \ref{flying focus concept}, for a positively chirped pulse, the longest wavelength focuses first at the near edge of the focal region and the shortest wavelength last at the far edge. Consequently, the peak intensity co-propagates forward at a subluminal speed. In contrast, applying a negative chirp reverses this temporal order: the shortest wavelength focuses first at the far edge, and longer wavelengths successively focus closer to the plate. This allows the focus to counter-propagate at a superluminal velocity. By precisely engineering the chirp parameter, the required focal velocities for various applications can be seamlessly achieved.

\subsection{Design of DOEs for flying focus}
%=======================================
%=======================================
\begin{figure}
    \centering
    \includegraphics[width=0.7\linewidth]{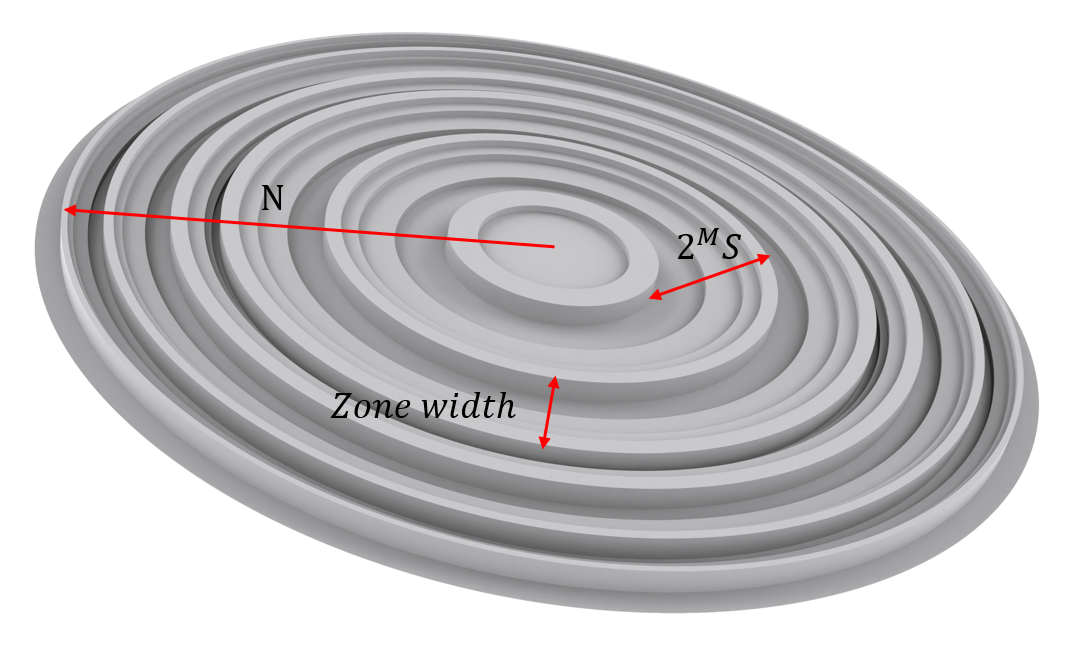}
    \caption{Schematic illustration of the optimized DOE design parameters. The optimization directions include the random modulation of zone widths (denoted as Zone width), the discretization of zone phase into multiple steps (indicated as \(2^M S\)), and the total number of zones \(N\).The stepwise phase profile is physically realized via surface relief, where each discrete phase step corresponds to a specific etch depth. The detailed conversion between the designed phase and the actual etch depth is provided in Appendix A.}
    \label{fig:doe_design}
\end{figure}

To achieve the aforementioned flying focus properties, we systematically optimize the DOEs' design through three key degrees of freedom as illustrated in Fig.~\ref{fig:doe_design}: the random modulation of zone widths (A parameter), the multi-level discrete phase steps (B parameter), the total number of zones \(N\) (C parameter). Then we also tune the chirp of the incident pulse (D parameter). While FZPs suffer from unwanted higher-order foci due to their strict radial periodicity, our design introduces random width modulation. This statistical strategy mimics the continuous transmittance of a GZP, and the resulting statistical equivalence is the key to eliminating higher-order foci. Furthermore, to address the efficiency limitation of binary amplitude masks, we introduce multi-level discrete phase modulation. By converting previously opaque zones into specific phase-shifting regions, this approach enhances constructive interference at the primary focal spot, thereby boosting focusing efficiency and further suppressing higher-order diffraction. In addition, a sufficiently large number of zones \(N\) is crucial to ensure a large numerical aperture (NA) and a small f-number, which provides an additional degree of suppression for higher-order foci and modify the size of focus spot at the same time. Finally, by tuning the chirp of the incident Gaussian pulse, the focal velocity is decoupled from the group velocity. This enables the flying focus to propagate at arbitrary velocities and directions.

% To achieve the aforementioned flying focus properties, we systematically optimize the zone plate structures and the parameter of the light. Zone plates control incident electric fields (\(E=Ae^{i\phi}\)) by modulating either the amplitude or the phase. FZPs act as binary amplitude masks and suffer from unwanted higher-order foci, which have equal amplitude of the main focus when illuminated by plane waves. The MWZPs overcome this by randomly modulating the transparent zone widths to mimic binary GZPs. However, functioning purely as an amplitude mask, the MWZP blocks the opaque zones, leading to low focusing efficiency around 10\%. To address this limitation, our MWZPs with multiple steps introduce phase modulation (\(e^{i\theta}\)) as a new degree of freedom. Instead of directly blocking light with opaque rings, it divides each zone and imparts specific phase delays. This phase-modulated design not only refines the single-focus characteristic but also significantly improves the focusing efficiency by utilizing the previously blocked light. Besides the introduction of multisteps,
% enough number of zones is also crucial for the zone plate to havs a sufficiently large NA and
% small f-number, which further suppress all higher-order foci. After optimizing the zone plate, the incident chirped Gaussian pulse is also needed to carefully consider. It is simulated by the modified Fresnel integrals \cite{PhysRevA.97.033835} and its chirp is another important parameter to tune.

\begingroup
\renewcommand{\thesubsubsection}{\Alph{subsubsection}} % 将编号改为大写字母
\setcounter{subsubsection}{0} % 重置计数器，确保从 A 开始

\subsubsection{Modulating zone width randomly (MWZP)}
% An MWZP is constructed by changing the width of every Fresnel rings \cite{ZHANG2018387,Li:26}: each continuous-transmission zone pair of a Gabor zone plate is replaced by a binary pair centered at \(R_{2n-1}\), and the transparent-ring width is sinusoidally modulated. 

% 理论开始
% \subsection{Statistical design of MWZP via Monte Carlo sampling}
% \label{sec:mwzp_design}

The fundamental challenge in designing a binary zone plate with a single focus lies in mimicking the continuous sinusoidal transmittance of a GZP while maintaining a step-like profile that is compatible with standard nanofabrication. To address this, the MWZP is introduced by randomizing the widths of the transparent zones. 

Following the design method of MWZP, the center of each zone pair is fixed at \(R_{2n-1}\).
The width of the \(n\)th transparent ring is\cite{ZHANG2018387},
\begin{equation}
d_n = 2(R_{2n}-R_{2n-1})s_n,
\label{eq:mwzp_dn}
\end{equation}
where \(R_m\) are the Fresnel-zone radii and \(s_n\) is a random variable with probability density\cite{ZHANG2018387},
\begin{equation}
\rho(s) = 0.5\pi \cos\!\left[\pi(s-0.5)\right], \quad 0 \le s \le 1.
\label{eq:mwzp_pdf}
\end{equation}
The cumulative distribution function F(s) is then obtained by integrating the probability density function $\rho(\mathrm{s})$\cite{Feller1971Probability}, 
%according to the standard mathematical procedure,
\begin{equation}
F(s) = \int_{0}^{s}\rho(s)ds=0.5 \sin \left[ \pi (s - 0.5) \right] + 0.5.
\end{equation} 
Then $\mathrm{s}$ can be obtained inversely as,
 \begin{equation}\label{eq.sin.random}
s = F^{-1}(s)= 0.5 + \frac{1}{\pi} \arcsin(2u - 1),
\end{equation}
where u is between 0 and 1. 
The resulting MWZP transmittance is \cite{ZHANG2018387},
\begin{equation}
t_{\mathrm{MWZP}}(r)=
\begin{cases}
1, & R_{2n-1}-\dfrac{d_n}{2} \le r < R_{2n-1}+\dfrac{d_n}{2},\\[4pt]
0, & \text{otherwise},
\end{cases}
\label{eq:mwzp_trans}
\end{equation}
with integer \(n\). 
% The cumulative distribution and inverse sampling of \(s_n\) are given in Ref.~\cite{Li:26}.

To theoretically demonstrate why this Monte Carlo random sampling suppresses higher-order foci, we examine the transmission probability at an arbitrary radius \(r\) within the \(n^{\mathrm{th}}\) zone pair. Transmission occurs when \(r\) falls inside the transparent ring, i.e., \(|r - R_{2n-1}| < d_n/2\). By defining a normalized relative distance \(\delta = |r - R_{2n-1}| / (R_{2n} - R_{2n-1})\), the condition becomes \(s_n > \delta\). The probability of transmission at radius \(r\) is therefore \(P(r) = 1 - F(\delta)\), and its expectation value over the random distribution is:
\begin{equation}
E[P(r)] = 1 - 0.5 \sin[\pi(\delta - 0.5)] - 0.5 = 0.5[1 + \cos(\pi \delta)].
\end{equation}
For a sufficiently large number of zone pairs, the local relative distance satisfies \(\pi \delta \approx \pi (r^2/R_1^2 - (2n-1))\). Detailed calculation about this approximation are illustrated in Appendix C. Consequently, the expected transmittance smoothly converges to the continuous transmittance function of a negative GZP, and this statistical equivalence is the key to eliminating higher-order foci, as the GZP inherently possesses no higher-order foci:
% Consequently, the expected transmittance smoothly converges to the continuous transmittance function of a negative GZP:
\begin{equation}
E[t_{\mathrm{MWZP}}(r)] = E[P(r)] \approx t_{\mathrm{GZP}}(r) = 0.5\left[1 - \cos\left(\pi \frac{r^2}{R_1^2}\right)\right].
\label{approxGZP}
\end{equation}

% Physically, this statistical equivalence is the key to eliminating higher-order foci, as the GZP has no high-order foci. 
Speaking with a more physical picture, the multiple foci of a standard FZP arise from the strict radial periodicity of its binary transmittance. For the \(m^{\mathrm{th}}\)-order focus, the phase difference between adjacent transparent zones is exactly \(2m\pi\). However, in the MWZP, the Monte Carlo sampling of \(s_n\) breaks this long-range strict periodicity. For the primary focus (\(m=1\)), the phase accumulation from the zone centers remains unperturbed, allowing constructive interference to be preserved. Conversely, for higher-order foci (\(m=3, 5, 7, \dots\)), the randomized widths introduce pseudo-random phase fluctuations among the diffraction waves from different zones. According to the Fresnel diffraction integral, the sum of these randomly phase-shifted contributions leads to destructive interference, effectively suppressing the higher-order intensity peaks and leaving only a single primary focus.

\subsubsection{Modulating the zone phase in multiple discrete steps (MWZP-$2^{M}$S)}
\begin{figure}
    \centering
    \includegraphics[width=1\linewidth]{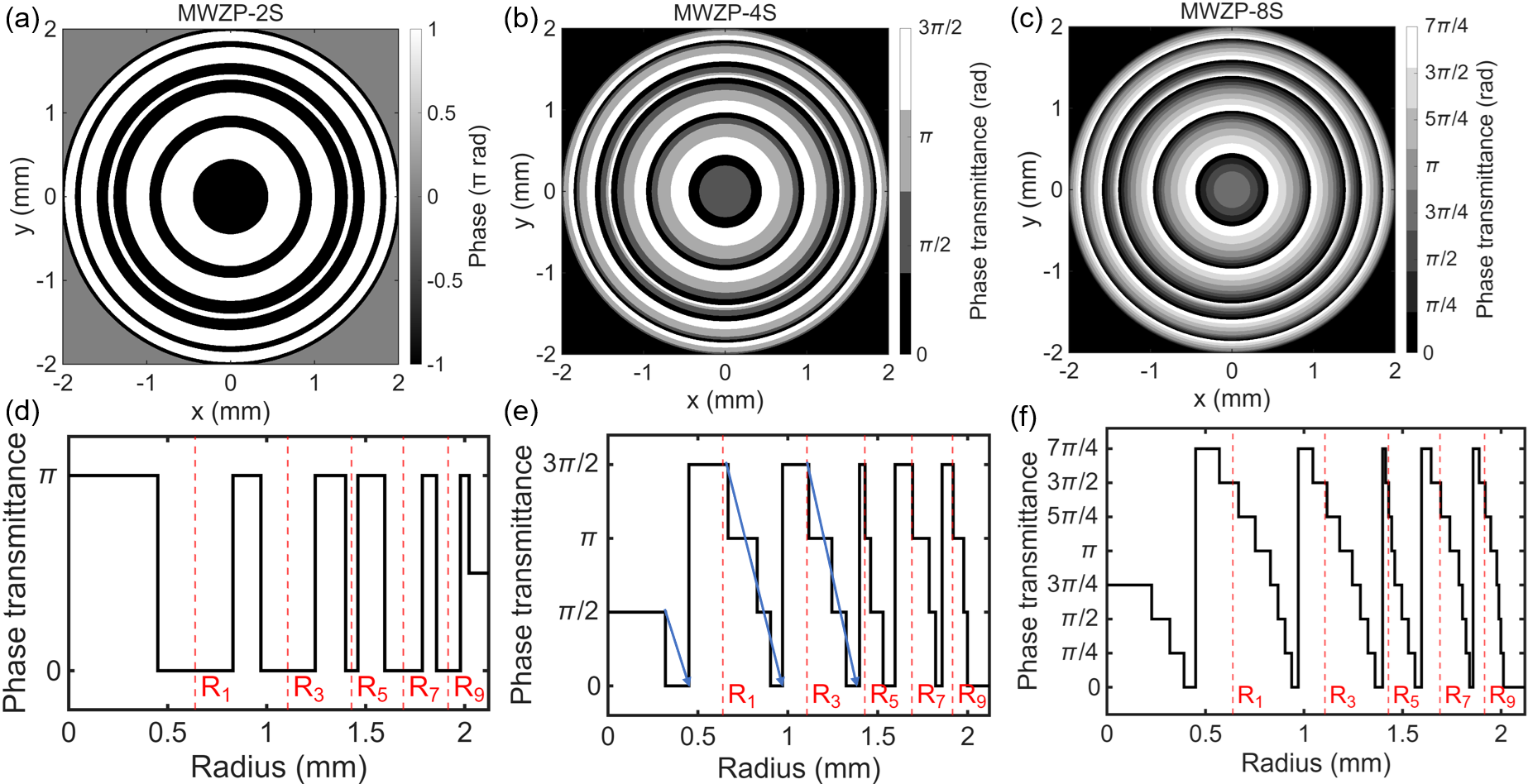}
    \caption{Phase distributions and transmittance characteristics of MWZP-\(2^M\)S (\(M=1, 2, 3\)). (a)-(c) Two-dimensional phase profiles of the MWZP-2S, MWZP-4S, and MWZP-8S (each with 10 zones), corresponding to \(M=1, 2, 3\), respectively. (d)-(f) Phase transmittance curves as a function of radius for the corresponding devices\cite{ZHANG2018387}.}
    \label{zone plate design}
\end{figure}

Zone plates control the incident electric field (\(E = Ae^{i\phi}\)) by modulating either the amplitude or the phase. To address the efficiency limitation of pure amplitude masks, our MWZPs with multiple steps introduce phase modulation (\(e^{i\theta}\)) as an additional degree of freedom. This phase modulation brings two major benefits: firstly, it significantly improves the focusing efficiency by converting the previously opaque zones into constructive phase-shifting zones; secondly, it further suppresses higher-order foci. 
% As demonstrated in later simulations, under the same number of zones, the relative intensity of high-order foci in the MWZP-8S is an order of magnitude smaller than that in the MWZP-4S. 

The primary goal of this multilevel phase quantization is to approach an ideal continuous Kinoform profile. As discussed in the review of diffractive optical elements \cite{10.3788/PI.2023.R09}, increasing the number of phase levels within a diffractive element allows the discrete phase profile to more closely approximate the ideal continuous phase, thereby concentrating the diffracted energy into the desired diffraction order. 
To achieve a uniform optical path difference (OPD) for each phase step, it is preferable to divide the subzones with unequal radial widths. Specifically, their boundaries follow the radius law of a multilevel FZP\cite{DiFabrizio1999}: 
\begin{equation}
r_{n,l} = \sqrt{\lambda f (2l/L + n - 2)}
\label{eq:r_nl}
\end{equation}
% \(r_{n,l} = \sqrt{\lambda f (2l/L + n - 2)}\)
, where \(L\) is the number of steps within each zone and \(l \in [1, L]\).
% Crucially, to ensure that each phase step introduces an identical optical path difference (OPD), the subzones must not be divided by equal radial width. Instead, their boundaries follow the radius law of a multilevel Fresnel zone plate (FZP)\cite{DiFabrizio1999}: \(r_{n,l} = \sqrt{\lambda f (2l/L + n - 2)}\), where \(L\) is the number of steps within each zone and \(l \in [1, L]\). 
Because the lens phase profile is proportional to the square of the radius (\(r^2\)), equal intervals in \(r^2\) guarantee a uniform phase increment across each subzone. This precise quantization improves constructive interference at the primary focal spot and effectively suppresses higher-order diffraction. The resulting phase distributions and corresponding transmittance curves for \(M=1, 2, 3\) are shown in Fig.~\ref{zone plate design}.

For the simplest case of \(M=1\), the MWZP-2S transforms the binary amplitude MWZP into a binary phase element. As illustrated in Fig.~\ref{zone plate design}(a) and (d), the transparent rings remain unchanged (transmission \(+1\)), while the originally opaque rings are replaced by zones that have a \(\pi\) phase shift (transmission \(-1\)). The resulting transmission function is purely binary in phase: \(t_{\mathrm{MWZP-2S}}(r) = 2t_{\mathrm{MWZP}}(r) - 1\). Because the energy that would otherwise be blocked by the opaque rings is now redirected constructively to the primary focus, the focal peak intensity is enhanced by a factor of 4 compared to that of the binary MWZP. 
% For \(M=2\), the MWZP-4S divides each zone pair into four \(r^2\)-equal subzones with phase delays of \([0.5, 0, 1.5, 1.0]\pi\), as shown in Fig.~\ref{zone plate design}(b) and (e). 

For \(M=2\), the MWZP-4S divides each zone pair into four \(r^2\)-equal subzones. The phase delays from the inner to the outer subzones follow a cyclic gradient descent with a step of  (\(\pi/2\)). Specifically, the innermost annular region consists of a half-cycle gradient (\( 0.50\pi, 0\)) and the phase sequence follows \([1.50, 1.00, 0.50, 0]\pi\), forming repeating complete cycles of the \(\pi/2\) downward step, as shown in Fig.~\ref{zone plate design}(b) and (e).

The MWZP-8S (\(M=3\)) is a key design in this work. Because the number of phase steps is greater, its capability to suppress higher-order foci is significantly stronger. As depicted in Fig.~\ref{zone plate design}(c) and (f), the phase delays from the inner to the outer subzones follow a cyclic gradient descent with a step of  (\(\pi/4\)). 
% Specifically, the phase sequence is exactly \([0.75, 0.50, 0.25, 0, 1.75, 1.50, 1.25, 1.00]\pi\). 
The innermost annular region consists of a half-cycle gradient (\(0.75\pi, 0.50\pi, 0.25\pi, 0\)), and moving outwards, the phase delays follow repeating complete cycles of the \(\pi/4\) downward step. Due to this fine phase matching, the diffracted light energy is concentrated into the first diffraction order. The MWZP-8S can push the focal peak intensity to a magnitude higher compared to that of the MWZP-4S, and demonstrating greater energy-gathering capability.

\subsubsection{Tuning the number of zone plates}
In zone plate design, the zone count ($\mathrm{N}$) cannot be treated as an independent parameter, as it is inherently coupled to the system aperture and resolving capability. With the operating wavelength ($\mathrm{\lambda}$) and first-order focal length ($\mathrm{f}$) predefined, the radius of the outermost zone ($\mathrm{R}_\mathrm{N}$) sets the numerical aperture $\mathrm{NA} \approx \mathrm{R}_\mathrm{N}/\mathrm{f}$ (valid for $\mathrm{f} \gg \mathrm{R}_\mathrm{N}$), the f-number $\mathrm{f}^{\#} \approx \mathrm{f}/(2\mathrm{R}_\mathrm{N})$, and the minimal outermost zone width $\Delta \mathrm{R}_\mathrm{N} \approx \mathrm{\lambda}\mathrm{f}/(2\mathrm{R}_\mathrm{N}) = \mathrm{\lambda}/(2\cdot\mathrm{NA})$. This width dictates both fabrication feasibility and the diffraction-limited spatial resolution. From Eq.\ref{Rm}, the total zone number can be derived as $\mathrm{N} \approx \mathrm{R}_\mathrm{N}^2 / (\mathrm{\lambda} \mathrm{f})$. Consequently, adjusting $\mathrm{N}$ within our numerical simulations amounts to a systematic scaling of the aperture dimension $\mathrm{R}_\mathrm{N}$, which concurrently modifies $\mathrm{NA}$, $\mathrm{f}^{\#}$, and $\Delta \mathrm{R}_\mathrm{N}$.

To illustrate more explicitly, tuning \(\mathrm{N}\) provides a powerful degree of freedom to optimize the flying focus performance in two critical aspects. First, the suppression of higher-order foci fundamentally relies on having a sufficiently large \(\mathrm{NA}\) and a small f-number. Specifically, a larger \(\mathrm{N}\) leads to a larger \(\mathrm{NA}\), which dictates that the zone plate must be equipped with enough zones to statistically emulate the Gabor zone plate and effectively suppress all higher-order foci by at least two orders of magnitude. Second, the total zone number directly governs the minimum focal spot size of the flying focus. As described in  Section 2.1, throughout the extended focal region, the flying focus maintains a self-similar intensity profile with a minimum spot size \(w_0 \simeq \lambda_0 f_0 / (\pi R)\). Here, the incident beam radius \(R\) is practically constrained by the outermost zone radius \(\mathrm{R}_\mathrm{N}\) of the zone plate. By substituting \(R \approx \mathrm{R}_\mathrm{N} = \sqrt{\mathrm{N}\lambda_0 f_0}\) into the expression for \(w_0\), we obtain:
\begin{equation}
w_0 \simeq \frac{\lambda_0 f_0}{\pi \sqrt{\mathrm{N}\lambda_0 f_0}} = \frac{1}{\pi}\sqrt{\frac{\lambda_0 f_0}{\mathrm{N}}}.
\end{equation}
This straightforward mathematical relation indicates that the minimum spot size \(w_0 \propto 1/\sqrt{\mathrm{N}}\). Therefore, increasing the number of zones \(\mathrm{N}\) directly reduces the minimum focal spot size \(w_0\), providing a straightforward approach to achieve a tighter focus. 

\subsubsection{Modulating the chirp of incident Gaussian pulses}

To numerically design the corresponding optical elements, we employ the modified Fresnel integrals \cite{PhysRevA.97.033835}. Using the convolution theorem, the modified Fresnel integrals are expressed as a convolution of the Gaussian beam transmitted through the zone plate with a modified impulse response function. The transverse electric field of the chirped Gaussian pulse is then obtained directly from this convolution; the detailed expressions of the modified impulse response function are provided in Appendix B. This approach efficiently captures the spatiotemporal couplings required for accurate flying focus simulations.

Tuning the chirp of the incoming laser pulse directly controls the flying focus velocity, allowing arbitrary co- or counter-propagation across the extended focal range (Fig. \ref{fig:chirp_velocity}). The focal velocity \(v_f\), i.e., the propagation speed of the peak-intensity point, is given by \cite{PhysRevA.97.033835},
\begin{equation}
\frac{v_f}{c} = \left( \frac{2f}{\eta c \tau^2 \omega_0} \right)
\left( 1 + \frac{2f}{\eta c \tau^2 \omega_0} \right)^{-1},
\label{chirp velocity equation}
\end{equation}
where \(\tau\) is the transform-limited incident pulse duration, \(\eta\) is the chirp parameter, and \(f\) is the focal length of the zone plate. According to Eq.\ref{chirp velocity equation}, the focus velocity is subluminal for positive chirp and can surpass the vacuum light speed for negative chirp.

\endgroup

% \subsubsection{The design's objective for flying focus scheme}

% This work employs modified Fresnel integrals for the design of customized zone plates, aiming to produce a flying focus characterized by \emph{no high-order foci with reduced background noise} and \emph{controllable focus velocities across the extended focal range}. 

%  To verify the design objectives, both a chirped Gaussian pulse and a monochromatic plane wave are used as the input beams. Three types of zone plates—FZPs, MWZPs, and multilevel Phase MWZPs—are included in the propagation process.

% Applying the modified Fresnel integrals to designed zone plates, this study aims to generate the flying focus with  \emph{no high-order foci with reduced background noise} and \emph{controllable focus velocities across the extended focal range}.
% A schematic of the setup is presented in Figure \ref{flying focus concept}, 
% which illustrates that a chirped Gaussian pulse passes through a Four-zone Phase MWZP to generate a lengthened focal region. 
 
%  The intensity distributions on the output planes are simulated using the modified Fresnel integrals for the chirped pulse and the standard angular spectrum method \cite{Goodman1969} for the plane wave. 

% The chirped Gaussian pulse, as well as the plane wave will be used as the input.

%==============================================================
%==============================================================
%==============================================================
%==============================================================
\section{Numerical Simulation}

Unless otherwise specified, the following parameters are used in the simulation.
The incoming laser has a central wavelength of 800 nm,  pulse duration of 115 fs, peak power of 16.3 kW, and chirp parameters of 109.1 and -220.0.
For the zone plates simulated here, 
multiple types of zone plates, including FZP, MWZP, and MWZP-$2^{M}$S,
%as in Figure \ref{zone plate design}) 
are considered, which all have primary focal lengths of 51 cm.
The input planes, output planes, and zone plates are all discretized into $7800\times 7800$ pixels.
%{\lzk The pixel number in a single transverse dimension, 
The number 7800 is chosen to qualify the stringent standard $\frac{\mathrm{k}_0\mathrm{w}_0}{\pi \mathrm{f}^{\#}}$\cite{PhysRevA.97.033835}, where $\mathrm{f}^{\#}$ is the f-number of the zone plate considered.
Table \ref{parmeter table1}  lists the major parameters used in the numerical simulations.

The conventional angular spectrum method\cite{Goodman1969IntroductionTF}
is employed to simulate the  plane wave propagation through the zone plates;
while for { the chirped Gaussian pulse}, the modified Fresnel integrals illustrated in Appendix B \cite{PhysRevA.97.033835} is used.
The output beam intensity  is described by the function $\mathrm{I}_{\mathrm{N}}^{\mathrm{ZT}}(\mathrm{x},\mathrm{y},\mathrm{z},\mathrm{t})$, where the superscript $\mathrm{ZT}$ indexes the zone-plate type,
and the underscript $\mathrm{N}$ signifies the total number of zones used.
% {\lzk For different kind of zone plates, the MWZP-2S is signified as 2MWZP due to its two-step phase distribution; and the MWZP-4S is also signified as 4MWZP in figures' legends.}
By modulating zone plate width  $\mathrm{d}_\mathrm{n}$, the number of  discrete steps $2^{M}$S,
zone number N, 
and the chirp parameter  $\eta$, 
the corresponding output intensity $\mathrm{I}_{\mathrm{N}}^{\mathrm{ZT}}(\mathrm{x},\mathrm{y},\mathrm{z},\mathrm{t})$ can be obtained.

The relative intensity, normalized by the intensity of the main focus, $\mathrm{R}_{\mathrm{N}}^{\mathrm{ZT}}(z)$, can be defined as,
\begin{equation}\label{eq.Rn.zt}
    R_{N}^{ZT}(z)=\frac{I_{N}^{ZT}(x=0,y=0,z,t=\frac{z}{c})}{I_{N}^{ZT}(x=0,y=0,z=f_1,t=\frac{f_1}{c})},
\end{equation}
where $\mathrm{f}_1$ is the primary focal length.

In this work, to more realistically evaluate the energy-gathering capability akin to a camera aperture, the focusing efficiency \(\eta(R_{ap})\) is defined using the variable aperture concept. Considering the aperture gradually expanding at the focal plane, the efficiency is calculated as the ratio of the integrated intensity within an aperture of radius \(R_{ap}\) to the total intensity incident on the entire zone plate. It is given by:

\begin{equation}
\eta(R_{ap}) = \frac{\int_{0}^{2\pi} \int_{0}^{R_{ap}} \mathrm{I}_{\mathrm{N}}^{\mathrm{ZT}}(r, \theta, z=f, t) \, r \, dr \, d\theta}{\mathrm{E}_{\mathrm{inc}}}
\label{eq:efficiency}
\end{equation}

\noindent where \(f\) is the focal length, and \(\mathrm{E}_{\mathrm{inc}}\) denotes the total incident energy. To obtain the focusing efficiency, the sampling pixels is increased to be 44800 in each side.
The derivative of the focusing efficiency with respect to the aperture radius, \(d\eta/dR_{ap}\), describes the spatial distribution of the incremental energy collection and is mathematically expressed as:

\begin{equation}
\frac{d\eta(R_{ap})}{dR_{ap}} = \frac{R_{ap}}{\mathrm{E}_{\mathrm{inc}}} \int_{0}^{2\pi} \mathrm{I}_{\mathrm{N}}^{\mathrm{ZT}}(R_{ap}, \theta, z=f, t) \, d\theta
\label{eq:derivative}
\end{equation}

\noindent Assuming cylindrical symmetry, Eq.~\eqref{eq:derivative} simplifies to \(\frac{d\eta}{dR_{ap}} \propto R_{ap} \cdot \mathrm{I}_{\mathrm{N}}^{\mathrm{ZT}}(R_{ap})\). This derivative curve exhibits a distinct peak, which physically corresponds to the highest rate of energy accumulation. According to the Fresnel-Kirchhoff diffraction theory \cite{Born1980Principles}, the theoretical Full Width at Half Maximum (FWHM) is \(1.028\delta_{RN}\). More detailed derivations are shown in Appendix C.

\begin{table}[htbp]
\caption{Parameters used in numerical simulations.}
  \label{tab:shape-functions}
  \centering
\begin{tabular}{cccc}
\hline
Type& Parameters & Value \\
\hline
 Chirped  Gaussian Pulse parameters & wavelength $\lambda (nm)$ & 800  \\
& $\Delta\lambda (nm)$ & 19  \\
&$\tau(fs)$ & 115 \\
& $P_{peak}(kW)$ & 16.3 \\
&chirp parameter $\eta$& 109.1 and -220.0\\
%\hline
%&Zone plate parameters & Value \\
%\hline
Zone plate parameters &The first zone radius $\rho_1 (\mu m)$ & $638.75$  \\
&Focal length $f_1(cm)$ & $51$  \\
&Maximum considered $w_0$(cm) & $2.8$  \\
&Pixels in a transverse dimension  & $7800$  \\
\hline
\end{tabular}
\label{parmeter table1}
\end{table}

\subsection{Modulating zone width randomly}
%...................................................................................................
% comparison_3plates_revision

\begin{figure}[htbp]
\centering \includegraphics[width=1\linewidth]{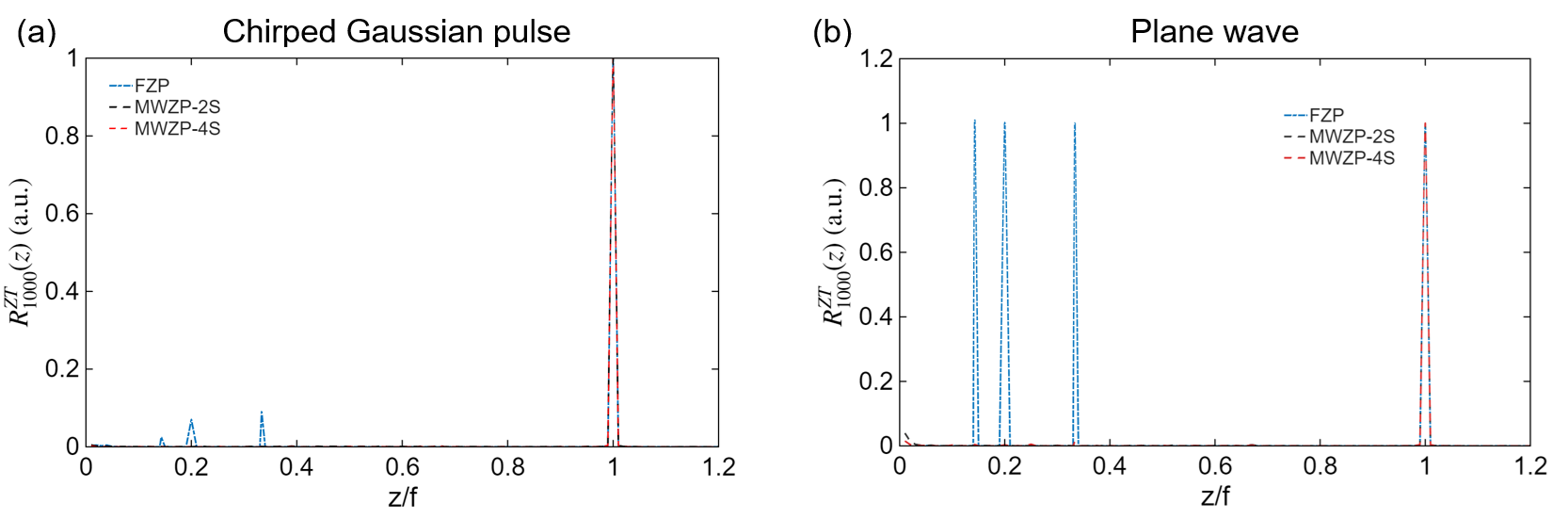}
\caption{Simulated on-axis normalized diffraction intensity distribution $\mathrm{R}_{1000}^{\mathrm{ZT}}(\mathrm{z})$ (in units of z/f) for (a) a chirped Gaussian pulse and (b) monochromatic plane wave propagation. The profiles compare the performance of an FZP with 1000 zones (blue dotted line) with a MWZP-2S (1000 zones, black dashed line) and an MWZP-4S (1000 zones, red dotted line).
%{\fcb （a/b/c）位置不统一}
}
\label{comparison_3plates_revision}
\end{figure}

By sinusoidally modulating zone widths,  a MWZP-2S is analyzed, together with a
MWZP-4S and an FZP.
Figure \ref{comparison_3plates_revision}a compares the on-axis normalized  intensity $\mathrm{R}_{1000}^{\mathrm{ZT}}(\mathrm{z}) $ of the FZP, MWZP-2S, and MWZP-4S  
 where the input beam is a chirped Gaussian pulse, described by Eq. \ref{input_beam}. 
 % Too short pulse width will not have enough time to generate the high-order foci.
 %\blue(“Gaussian” refers to the spatial intensity profile of the beam's transverse cross-section)
% \fcb 直接给出方程。
% which has a pulse form, .....
One can find that the ratios of the intensities of  the higher-order foci  and the main focus  intensities of the MWZP-2S and MWZP-4S are suppressed. MWZP-8S (not shown here) display similar suppression of higher-order foci.
However, for the FZP, the 3rd, 5th, and 7th-order foci still exist. Their relative low magnitude are due to the fs-level short pulse.

For comparison, 
if the input beam  is a monochromatic plane wave, 
instead of a chirped Gaussian pulse,
%which is defined as Eq.???
the results are shown in Figure \ref{comparison_3plates_revision}b. The results are similar as those in the chirped Gaussian pulse case.

\subsection{Modulating the zone phase in multiple discrete steps}

\begin{figure}[htbp]
\centering \includegraphics[width=1\linewidth]{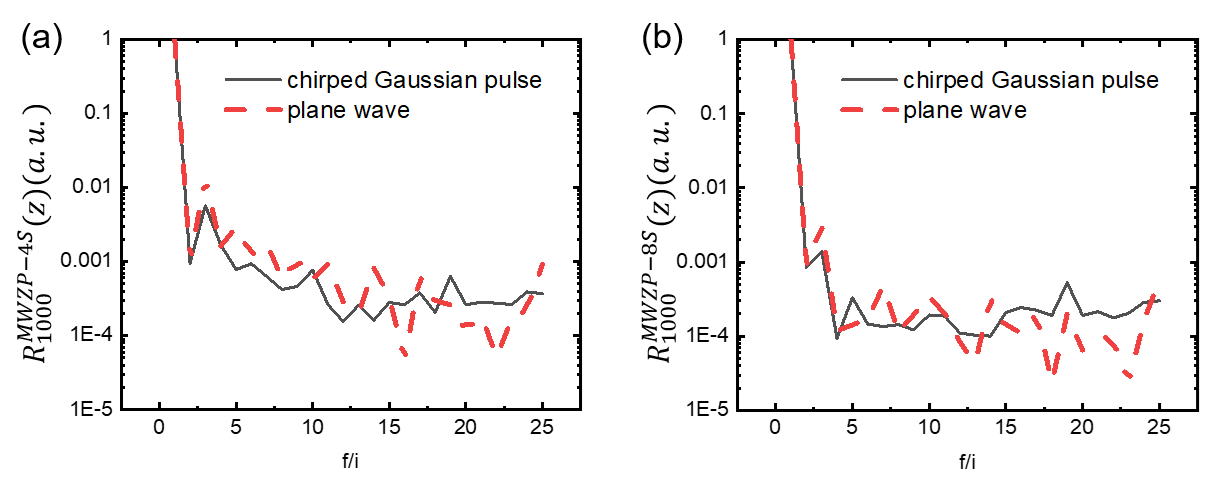}
\caption{Normalized intensities at discrete higher-order focal positions \(f/i\) (where \(i\) is an integer) for (a) the MWZP-4S (\(\mathrm{R}_{1000}^{\mathrm{MWZP-4S}}(\mathrm{z})\)) and (b) the MWZP-8S (\(\mathrm{R}_{1000}^{\mathrm{MWZP-8S}}(\mathrm{z})\)). Both propagation cases, a chirped Gaussian pulse (black solid line) and a plane wave (red dashed line), are distinctly shown for comparison.
}
\label{high_order_foci}
\end{figure}

Figures \ref{high_order_foci} illustrate the normalized intensities \(\mathrm{R}_{1000}^{\mathrm{MWZP-4S}}(z)\) and \(\mathrm{R}_{1000}^{\mathrm{MWZP-8S}}(z)\) at discrete higher-order focal positions \(f/i\) (where \(i\) is an integer), under both a chirped Gaussian pulse (black solid line) and a plane wave (red dashed line). To quantitatively evaluate the suppression of higher-order foci, we compare the relative intensity at the fifth-order focal spot (\(f/5\)) under plane wave illumination. The relative intensity is  \(2.9 \times 10^{-3}\) for the MWZP-4S, whereas it decreases to \(1.4 \times 10^{-4}\) for the MWZP-8S, representing a \(20\)-fold further suppression. Even for a chirped Gaussian pulse which involves multiple wavelengths, the relative intensities at higher-order focal spots in the MWZP-8S remain firmly depressed to the order of \(10^{-4}\), demonstrating that increasing the number of phase steps significantly enhances the single-focus capability.
% Figure \ref{comparison_3plates_revision}d illustrates the $\mathrm{R}_{1000}^{\mathrm{MWZP-8S}}(\mathrm{z})$, where the input beams are a plane wave (red dashed line) and a chirped Gaussian pulse (black solid line). 
% Even for a chirped Gaussian pulse which involves multiple wavelengths, 
% the relative intensities at  high-order focal spots can still be depressed more than 1000 times. }

%............................................................................
%............................................................................
%............................................................................

%............................................................................
%............................................................................
%............................................................................

\begin{figure}[htbp]
\centering \includegraphics[width=1\linewidth]{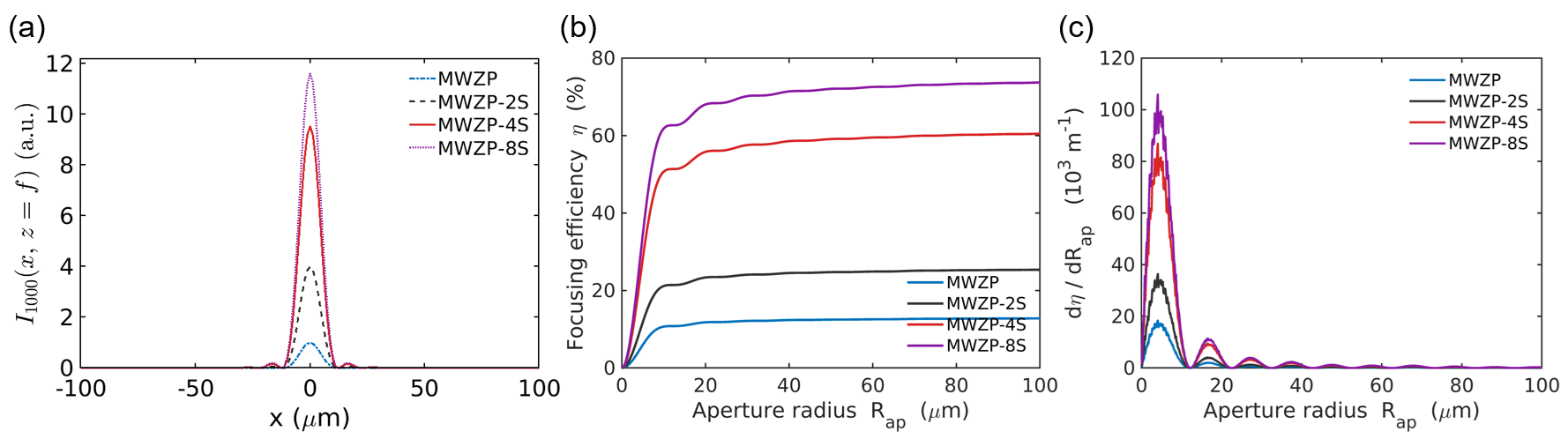}
\caption{(a) Simulated diffraction intensity distribution along the x axis on the focal plane. The blue, black, red, and purple lines represent the results of MWZP, MWZP-2S, MWZP-4S, and MWZP-8S, respectively, all with 1000 rings. The central diffraction intensity of MWZP is normalized to unity. (b) Focusing efficiency of MWZP, MWZP-2S, MWZP-4S, and MWZP-8S as a function of the aperture radius. (c) Derivative of focusing efficiency of MWZP, MWZP-2S, MWZP-4S, and MWZP-8S with respect to the aperture radius.
%{\fcb y-label: $I_{N}^{zt}(x=0,y=0,z,t) $}
}
\label{changing_plate_number2}
\end{figure}

% The full width at half maximum (FWHM) of the focal spot is equivalently determined at the reference radius \(R_{ap} = 1.028\delta \mathrm{R}_{\mathrm{N}}\), representing the optimal balance between the main lobe energy and resolution.

% The focusing efficiency is defined as the ratio of the integrated intensity of the focal spot to the total intensity incident on the entire zone plate. The focal spot region is determined using the intensity threshold method: taking the peak intensity at the focal center as the baseline, the contour where the intensity drops to {\lzk one half} of the peak intensity is used as the boundary, and the area enclosed within this contour is defined as the effective focal spot.

% Figure \ref{changing_plate_number2}(a) indicates that MWZP, MWZP-2S, and MWZP-4S achieve nearly identical focal spot sizes under the same simulation parameters. In our simulation, the outermost zone width is \(\delta_{RN} = 10.1020\,\mu\text{m}\), yielding a theoretical FWHM reference of \(1.028\delta_{RN} = 10.3849\,\mu\text{m}\). The measured FWHM values for MWZP,  MWZP-2S, and MWZP-4S are \(10.346\,\mu\text{m}\), \(10.346\,\mu\text{m}\), and \(10.371\,\mu\text{m}\), respectively. These values correspond to \(99.63\%\), \(99.63\%\), and \(99.87\%\) of the theoretical FWHM reference. Consequently, compared to the standard MWZP, the MWZP-2S and MWZP-4S do not weaken the capability in focusing light, as their focal spot sizes perfectly approach the diffraction limit.

Figure \ref{changing_plate_number2}(a) indicates that the MWZP, MWZP-2S, MWZP-4S, and MWZP-8S achieve nearly identical focal spot sizes under the same simulation parameters. In our simulation, the outermost zone width is \(\delta_{RN} \approx 10.1\,\mu\text{m}\), yielding a theoretical FWHM reference of approximately \(10.4\,\mu\text{m}\). The measured FWHM values for all four structures range from \(10.35\,\mu\text{m}\) to \(10.38\,\mu\text{m}\), showing excellent agreement with the theoretical value. Consequently, compared to the standard MWZP, the multi-step phase plates (MWZP-2S, MWZP-4S, and MWZP-8S) do not weaken the capability in focusing light; rather, their focal spot sizes perfectly approach the diffraction limit.

% Figure \ref{changing_plate_number2}(a) indicates that the MWZP, MWZP-2S, MWZP-4S, and MWZP-8S achieve nearly identical focal spot sizes under the same simulation parameters. In our simulation, the outermost zone width is \(\delta_{RN} = 10.1020\,\mu\text{m}\), yielding a theoretical FWHM reference of \(1.028\delta_{RN} = 10.3849\,\mu\text{m}\). The measured FWHM values for the MWZP, MWZP-2S, MWZP-4S, and MWZP-8S are \(10.346\,\mu\text{m}\), \(10.346\,\mu\text{m}\), \(10.370\,\mu\text{m}\), and \(10.376\,\mu\text{m}\), respectively. These values correspond to \(99.63\%\), \(99.63\%\), \(99.86\%\), and \(99.92\%\) of the theoretical FWHM reference. Consequently, compared to the standard MWZP, the multi-step phase plates (MWZP-2S, MWZP-4S, and MWZP-8S) do not weaken the capability in focusing light; rather, their focal spot sizes perfectly approach the diffraction limit.

\begin{table}[htbp]
\caption{Comparison of peak intensity and focusing efficiency at the FWHM radius (1.028\(\delta R_N\)) and the saturated aperture (100~\(\mu\)m) for different phase plates.}
\label{efficiency_comparison_table}
\centering
\begin{tabular}{lccc}
\hline
Phase plate & Peak intensity & Efficiency @ FWHM (\%) & Efficiency @ 100~\(\mu\)m (\%) \\
            & (relative to MWZP) & (1.028\(\delta R_N\)) & (Saturation) \\
\hline
MWZP    & 1.0  & 10.8 & 12.8 \\
MWZP-2S & 4.0  & 21.3 & 25.4 \\
MWZP-4S & 9.5  & 51.0 & 60.5 \\
MWZP-8S & 11.6 & 62.2 & 73.7 \\
\hline
\end{tabular}
\end{table}
% \begin{table}[htbp]
% \caption{Comparison of peak intensity and focusing efficiency at the FWHM radius (1.028\(\delta R_N\)) and the saturated aperture (100~\(\mu\)m) for different phase plates.}
% \label{tab:peak_intensity_efficiency}
% \centering
% \begin{tabular}{lccc}
% \hline
% Phase plate & Peak intensity & Efficiency @ FWHM (\%) & Efficiency @ 100~\(\mu\)m (\%) \\
%             & (relative to MWZP) & (1.028\(\delta R_N\)) & (Saturation) \\
% \hline
% MWZP  & 1.0000 & 10.75 & 12.81 \\
% MWZP-2S & 4.0000 & 21.29 & 25.37 \\
% MWZP-4S & 9.5380 & 51.01 & 60.49 \\
% \hline
% \end{tabular}
% \end{table}

% Figure \ref{changing_plate_number2}(b) illustrates the variation of focusing efficiency \(\eta\) with the aperture radius \(R_{ap}\) for the MWZP, MWZP-2S, and MWZP-4S. As \(R_{ap}\) increases, the focusing efficiencies of all three types of zone plates rise rapidly and eventually saturate, exhibiting distinct step-like features corresponding to the concentric rings of the focal spots. As shown in table \ref{tab:peak_intensity_efficiency}, taking the FWHM reference radius as a metric for the central lobe energy, the focusing efficiencies are 10.75\%, 21.29\%, and 51.01\% for MWZP, MWZP-2S, and MWZP-4S, respectively. When the aperture expands to 100~\(\mu\)m (approaching saturation), the ultimate energy collection efficiencies reach 12.81\%, 25.37\%, and 60.49\%, respectively. 

Figure \ref{changing_plate_number2}(b) illustrates the variation of focusing efficiency \(\eta\) with the aperture radius \(R_{ap}\) for the MWZP, MWZP-2S, MWZP-4S, and MWZP-8S. As \(R_{ap}\) increases, the focusing efficiencies of all four types of zone plates rise rapidly and eventually saturate, exhibiting distinct step-like features corresponding to the concentric rings of the focal spots. As shown in Table \ref{efficiency_comparison_table}, taking the FWHM reference radius as a metric for the central lobe energy, the focusing efficiencies are 10.8\%, 21.23\%, 51.0\%, and 62.2\% for MWZP, MWZP-2S, MWZP-4S, and MWZP-8S, respectively. When the aperture expands to 100~\(\mu\)m (approaching saturation), the ultimate energy collection efficiencies reach 12.8\%, 25.4\%, 60.5\%, and 73.7\%, respectively.

% The binary amplitude nature of the MWZP, where half of its zones are opaque, fundamentally limits its efficiency, resulting in severe energy loss and making it unsuitable for high-intensity laser experiments. The MWZP-2S evolves by replacing the opaque regions with \(\pi\)-phase-shifting zones. The constructive interference between the \(\pi\)-phase and \(0\)-phase zones optimally redistributes the energy, yielding a 4-fold enhancement in focal peak intensity compared to the MWZP.
% Furthermore, benefiting from its finer four-step phase segmentation, the MWZP-4S achieves a much higher focal peak intensity, which is 9.54 times that of the MWZP, and its FWHM efficiency is enhanced by 4.7 times. This continuous enhancement in both peak intensity and central energy concentration is visually consistent with the absolute intensity profiles shown in Fig.\ref{changing_plate_number2}(a).
The binary amplitude nature of the MWZP, where half of its zones are opaque, fundamentally limits its efficiency, resulting in severe energy loss and making it unsuitable for high-intensity laser experiments. The MWZP-2S evolves by replacing the opaque regions with \(\pi\)-phase-shifting zones. The constructive interference between the \(\pi\)-phase and \(0\)-phase zones optimally redistributes the energy, yielding a 4-fold enhancement in focal peak intensity compared to the MWZP. Furthermore, benefiting from its finer four-step phase segmentation, the MWZP-4S achieves a much higher focal peak intensity, which is 9.5 times that of the MWZP, and its FWHM energy collection efficiency is enhanced by 4.7 times. Extending this multilevel strategy to eight steps, the MWZP-8S pushes the focal peak intensity to 12 times and the FWHM efficiency to 5.8 times that of the MWZP. This continuous enhancement in both peak intensity and central energy concentration is visually consistent with the absolute intensity profiles shown in Fig.\ref{changing_plate_number2}(a).

To further reveal the energy accumulation process, Fig.\ref{changing_plate_number2}(c) presents the derivative of the focusing efficiency with respect to the aperture radius, \(d\eta/dR_{ap}\), which physically represents the incremental energy collection rate. Consequently, the prominent peaks in Fig.\ref{changing_plate_number2}(c) at \(R_{ap} \approx 4~\mu\)m indicate the maximum rate of energy collection within the primary focal lobe, while the subsequent oscillations correspond to the concentric side lobes of the Airy-like patterns. As \(R_{ap}\) further increases, the derivative gradually approaches zero, confirming the complete collection of the focal spot energy and the saturation of the total efficiency. Here, MWZP-8S collects energy with the greatest derivative.

% The binary nature of the MWZP, where half of its zones are opaque, leads to low focusing efficiency. A significant amount of energy is lost during operation, making the MWZP unsuitable for high-intensity laser experiments requiring maximum focused intensity. The Phase MWZP evolves from the binary amplitude MWZP by replacing opaque regions with phase-shifting zones. The zones with transmission \(-1\) constructively interfere with those of transmission \(+1\), yielding a twofold enhancement in focusing efficiency. Last but not least, benefiting from its finer phase segmentation, the Four-zone Phase MWZP achieves further improved focusing efficiency. Compared with the MWZP, its focusing efficiency is enhanced by an order of magnitude. This efficiency comparison is also presented in Fig.\ref{changing_plate_number2}(a), with consistent relative focusing intensities.

% However, the average focusing efficiency of MWZP is only 8.0{\%}, suggesting that a significant amount of energy is lost during its operation. Consequently, MWZP is not suitable for high-intensity laser experiments that demand the highest focusing intensity; nevertheless, due to its relatively simple fabrication process, it still holds certain application value.

\subsection{Tuning the number of zone plates}
% In zone plate design, the zone count ($\mathrm{N}$) cannot be treated as an independent parameter, as it is inherently coupled to the system aperture and resolving capability. With the operating wavelength ($\mathrm{\lambda}$) and first-order focal length ($\mathrm{f}$) predefined, the radius of the outermost zone ($\mathrm{R}_\mathrm{N}$) sets the numerical aperture $\mathrm{NA}$, f-number $\mathrm{f}^{\#}$, and the minimal outermost zone width \(\delta \mathrm{R_{N}}\). This width dictates both fabrication feasibility and the diffraction-limited spatial resolution. From Eq.\ref{Rm}, the total zone number can be derived as $\mathrm{N} \approx \mathrm{R}_\mathrm{N}^2 / (\mathrm{\lambda} \mathrm{f})$. Consequently, adjusting $\mathrm{N}$ within our numerical simulations amounts to a systematic scaling of the aperture dimension $\mathrm{R}_\mathrm{N}$, which concurrently modifies $\mathrm{NA}$, $\mathrm{f}^{\#}$, and \(\delta \mathrm{R_{N}}\).

\begin{figure}[htbp]
\centering \includegraphics[width=1\linewidth]{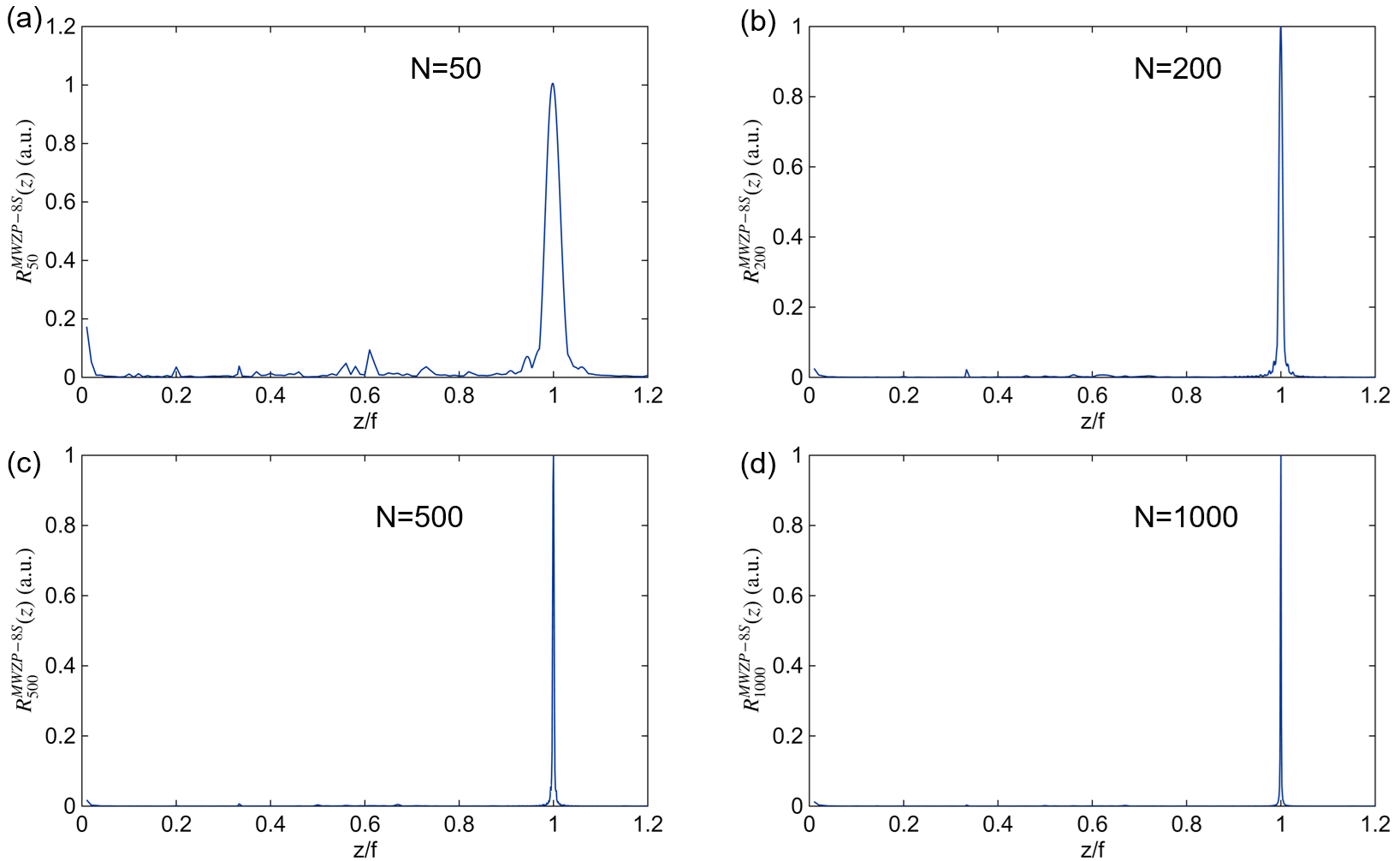}
\caption{Normalized diffraction intensity distribution $\mathrm{R}_{\mathrm{N}}^{\mathrm{MWZP-8S}}(\mathrm{z})$ along the optical axis for MWZP-8S with different number of zones\cite{ZHANG2018387}. (a) 50 zones. (b) 200 zones. (c) 500 zones. (d) 1000 zones. 
}
\label{multiple_500zone_pairs}
\end{figure}

Numerical simulations are performed to clarify the dependence of 
the  relative intensity $\mathrm{R}_{\mathrm{N}}^{\mathrm{MWZP-8S}}(\mathrm{z})$ (Eq.\ref{eq.Rn.zt}) and  the FWHM of the primary focal peak on the number of zones $\mathrm{N}$. 

% As shown in Figure \ref{multiple_500zone_pairs}, 
% the relative intensities $\mathrm{R}_{\mathrm{N}}^{\mathrm{4MWZP}}(\mathrm{z})$ for Four-zone Phase MWZP with different zone numbers  ($\mathrm{N} = 50, 200, 500$, and 1000) are presented. 
% When the MWZP has only N = 50 zones (Figure \ref{multiple_500zone_pairs}a), 
% the maximum relative intensity $\mathrm{R}_{\mathrm{N}}^{\mathrm{MWZP}}(\mathrm{z})$ has the order of 0.25 in the non-primary focal range.
% The intensity fluctuations  gradually decrease with increasing $\mathrm{N}$. 
% Meanwhile, %at the primary focal spot $z=f_1$, 
% the FWHM of the primary focal peak is reduced significantly,  from $\mathrm{FWHM}_{\mathrm{N} = 50} = 1.79$ cm to $\mathrm{FWHM}_{\mathrm{N} = 200} = 0.46$ cm, with increasing  N.

\begin{table}[htbp]
\caption{Maximum relative intensity in the non-primary focal range and the corresponding FWHM for different numbers of zone pairs.}
\label{tab:nonprimary_intensity_fwhm}
\centering
\begin{tabular}{ccc}
\hline
$N$ (zone pairs) & Maximum relative intensity $\mathrm{R}_{\mathrm{N}}^{\mathrm{MWZP-4S}}(\mathrm{z})$ & $\mathrm{FWHM}$ (cm) \\
\hline
50   & $\sim 9.4\%$   & 1.80 \\
200  & $\sim 2.2\%$   & 0.46 \\
500  & $\sim 0.6\%$  & 0.18 \\
1000 & $\sim 0.3\%$     & 0.10 \\
\hline
\end{tabular}
\end{table}

As shown in Figure \ref{multiple_500zone_pairs}, the relative intensities $\mathrm{R}_{\mathrm{N}}^{\mathrm{MWZP-4S}}(\mathrm{z})$  are presented for MWZP-4S with different zone numbers ($N = 50, 200, 500$, and $1000$). As $\mathrm{N}$ increases, both the maximum relative intensity in the non-primary focal range and the FWHM of the primary focal peak decrease monotonically; the specific values are summarized in Table \ref{tab:nonprimary_intensity_fwhm}. In particular, for the largest considered case of $N = 1000$ (Figure \ref{multiple_500zone_pairs}d), the maximum relative intensity is $<0.3\%$ and the FWHM is $0.10$ cm. These results demonstrate that the single-focus performance of the MWZP-8S becomes effective only when a sufficiently large $\mathrm{N}$ (i.e., a sufficiently large NA and small f-number) is achieved.

\begin{figure}[htbp]
\centering 
\includegraphics[width=1\linewidth]{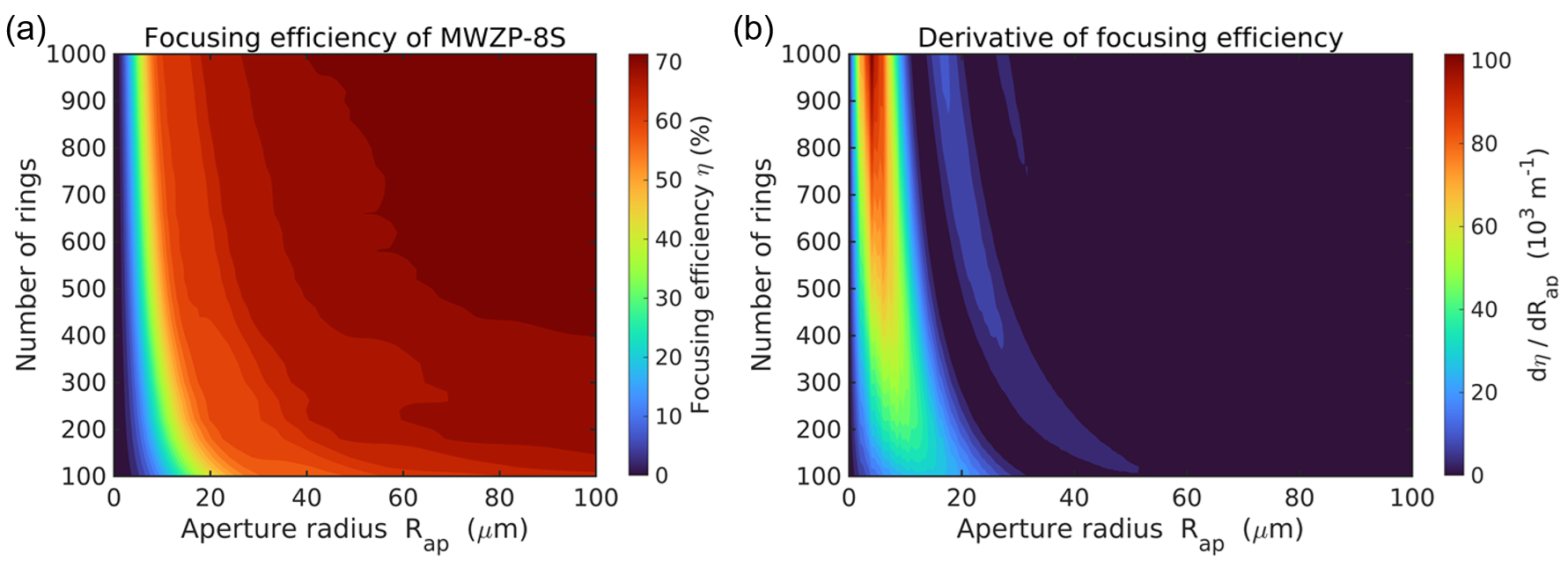}
\caption{The contour maps illustrating the evolution of focusing characteristics for the MWZP-8S as a function of the aperture radius \(R_{ap}\) and the number of rings. (a) The focusing efficiency \(\eta\) as a function of \(R_{ap}\) and the number of rings. The color bar represents the focusing efficiency in percentage. (b) The derivative of the focusing efficiency, \(d\eta/dR_{ap}\), as a function of \(R_{ap}\) and the number of rings. The prominent colored ridges in (b) represent the maximum rate of energy accumulation. Modulating the zone width randomly introduces some fluctuations to the efficiency's satuation ridge, but the derivative of focusing efficiency still display the physics clearly. }
\label{8MWZP_contour}
\end{figure}

Figure~\ref{8MWZP_contour} illustrates the evolution of focusing characteristics for the MWZP-8S as a function of the aperture radius \(R_{ap}\) and the number of rings. 
As shown in Fig.~\ref{8MWZP_contour}(a), the focusing efficiency saturates at a significantly smaller aperture as the number of rings increases, indicating that the incident energy is effectively concentrated into a tighter spatial region. This trend is further confirmed in Fig.~\ref{8MWZP_contour}(b), where the peak derivative ridges shift remarkably toward smaller aperture radii. 
While the random modulation of the zone widths introduces minor fluctuations into the saturation ridge of the focusing efficiency, the derivative map effectively filters out this interference and clearly delineates the underlying physical trend. The prominent colored ridges in Fig.~\ref{8MWZP_contour}(b) reliably track the maximum rate of energy accumulation despite the random structural variations.
In Section 2.2.C, increasing the number of zones \(\mathrm{N}\) directly reduces the minimum focal spot size \(w_0\)：\(w_0 \propto 1/\sqrt{\mathrm{N}}\). The colored ridges directly correspond to the energy distribution of the focal spot, which follows the previous scaling law.
These results demonstrate that plates with more zones possess stronger energy-concentration capabilities, which is highly advantageous for applications requiring focusing precision.

% For case of N = $50$  (Figure \ref{multiple_500zone_pairs}c), 
%  in the non-primary focal range, the maximum relative intensity $\mathrm{R}_{\mathrm{N}}^{\mathrm{4MWZP}}(\mathrm{z})$ is  approximately 25\%, and $\mathrm{FWHM}_{\mathrm{N} = 50} = 1.79$ cm; 

%  For case of N = $200$  (Figure \ref{multiple_500zone_pairs}c), 
%  in the non-primary focal range, the maximum relative intensity $\mathrm{R}_{\mathrm{N}}^{\mathrm{4MWZP}}(\mathrm{z})$ is  approximately 10\%, and $\mathrm{FWHM}_{\mathrm{N} = 200} = 0.46$ cm; 

% For case of N = $500$  (Figure \ref{multiple_500zone_pairs}c), 
%  in the non-primary focal range, the maximum relative intensity $\mathrm{R}_{\mathrm{N}}^{\mathrm{4MWZP}}(\mathrm{z})$ is  approximately 2.9\%, and $\mathrm{FWHM}_{\mathrm{N} = 500} = 0.18$ cm; 
%  and for case of N = 1000 (Figure \ref{multiple_500zone_pairs}d),
% they are $<1.4\%$ 
% and $\mathrm{FWHM}_{\mathrm{N} = 1000} = 0.10 $ cm  ,  correspondingly.
% Our results demonstrate that the single-focus performance of the Four-zone Phase MWZP becomes effective only when a sufficiently large N (i.e., a sufficiently large NA and small f-number) is achieved.

% 0.097 cm

\subsection{Modulating the chirp of incident pulses}

% As shown in Figure \ref{chirp velocity}, tuning the chirp of incoming laser pulses provides direct control over the flying focus velocity, enabling it to propagate or counter-propagate at arbitrary velocities across the extended focal range. 
% By using a Four-zone Phase MWZP with a radius of 2.8 cm, and 960 zone pairs,
% this effect is simulated.

% The focal velocity $\mathrm{v}_\mathrm{f}$ denotes the rate at which the point of peak intensity propagates, a continuous progression driven by the chromatic focusing and the temporal separation of frequencies within the pulse\cite{PhysRevA.97.033835}:
% \begin{equation}
% \frac{v_f}{c} = \left( \frac{2f}{\eta c \tau^2 \omega_0} \right) \left( 1 + \frac{2f}{\eta c \tau^2 \omega_0} \right)^{-1},
% \end{equation}
% where $\tau$ is the transform-limited incident pulse duration, $\eta$ is the chirp parameter, and f is the focus of the zone plate.

% In further specific simulations, the chirp parameter is chosen to be 3.1 and -220, marked by the blue circle and red square, respectively, in Figure \ref{chirp velocity}. These chirp values yield focus velocities close to the speed of light in the forward or backward directions.

\begin{figure}[htbp]
\centering 
\includegraphics[width=1\linewidth]{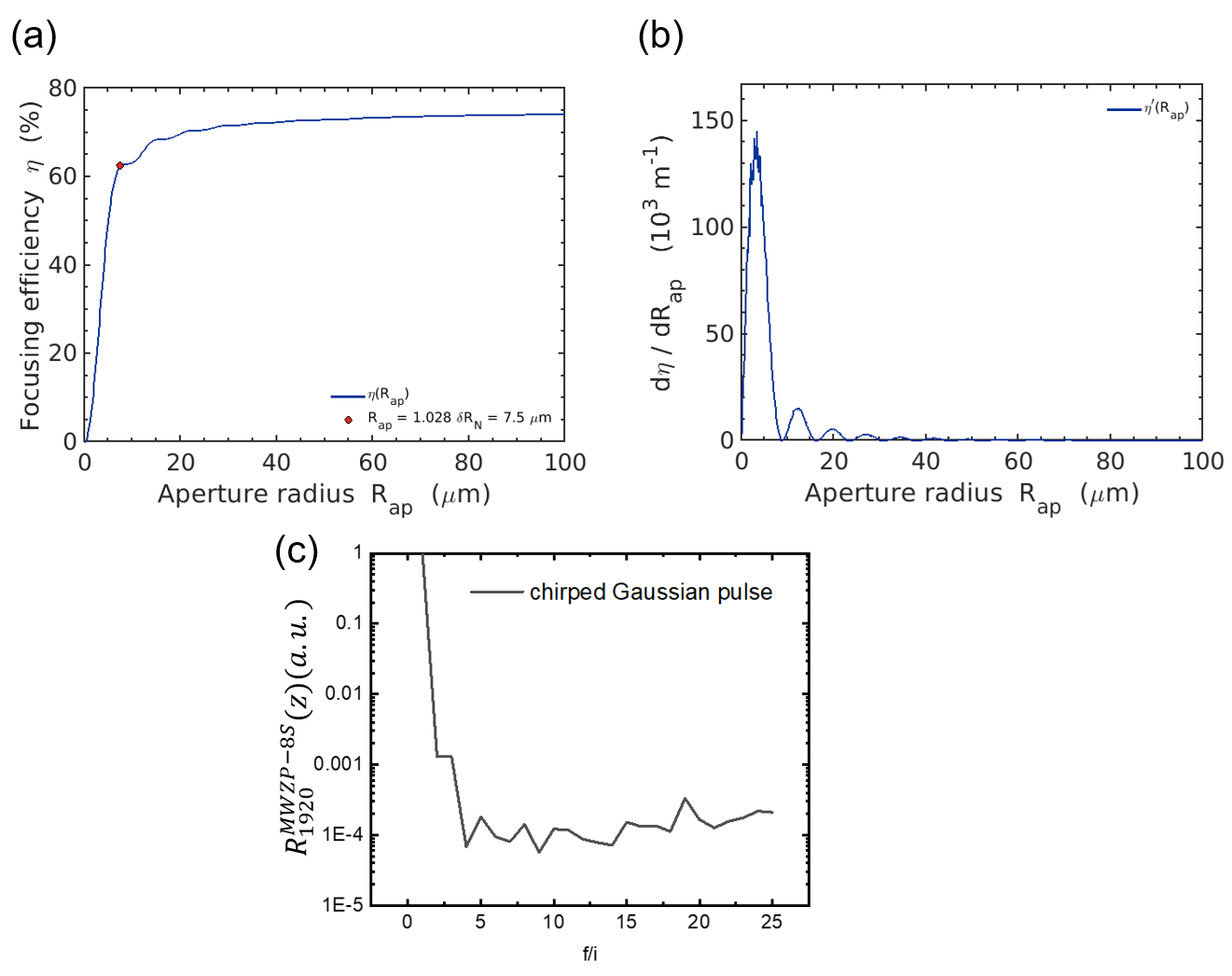}
\caption{(a) The focusing efficiency of the zone plate as a function of the aperture radius. The blue line represents the focusing efficiency curve, and the red dot marks the reference radius \(R_{ap} = 1.028 \delta R_N\). (b) The derivative of focusing efficiency as a function of the aperture radius. (c) Normalized intensities at discrete higher-order focal positions \(f/i\) (where \(i\) is an integer) for the MWZP-8S (\(\mathrm{R}_{1920}^{\mathrm{MWZP-8S}}(\mathrm{z})\)). A chirped Gaussian pulse with chirp parameter -220.0 is propagated. }
\label{efficiency_1920}
\end{figure}

To simulate the chirp effect, we employ a MWZP-8S with a radius of 2.8 cm and 1920 zones. As shown in Fig.\ref{efficiency_1920}(a), the FWHM of the focal spot is 7.5 \(\mu\)m. Such a compact focal spot is highly suitable for inverse Compton scattering with electron bunches of a similar transverse size \cite{gerstmayr2026experimentaldemonstrationflyingfocusenhanced}. At this reference radius, the focusing efficiency reaches 62.5\%, and the overall saturation efficiency approaches 74.3\%. Furthermore, compared to the previously discussed 1000-ring MWZP-8S, the derivative of the focusing efficiency exhibits a noticeably higher peak with greater focusing capability, as illustrated in Fig.~\ref{efficiency_1920}(b). 
In Fig.~\ref{efficiency_1920}(c), for a chirped Gaussian pulse with chirp -220.0, the relative intensities at higher-order focal spots in the MWZP-8S are depressed to the order of \(10^{-4}\).

Choosing \(\eta = 109.1\) and \(-220.0\) (blue circle and red square in Fig. \ref{fig:chirp_velocity}) yields focus velocities half of light speed in the forward direction and the speed of light in the backward direction, respectively.

\subsubsection{Flying focus with a positive chirp}
\begin{figure}[htbp]
\centering \includegraphics[width=1\linewidth]{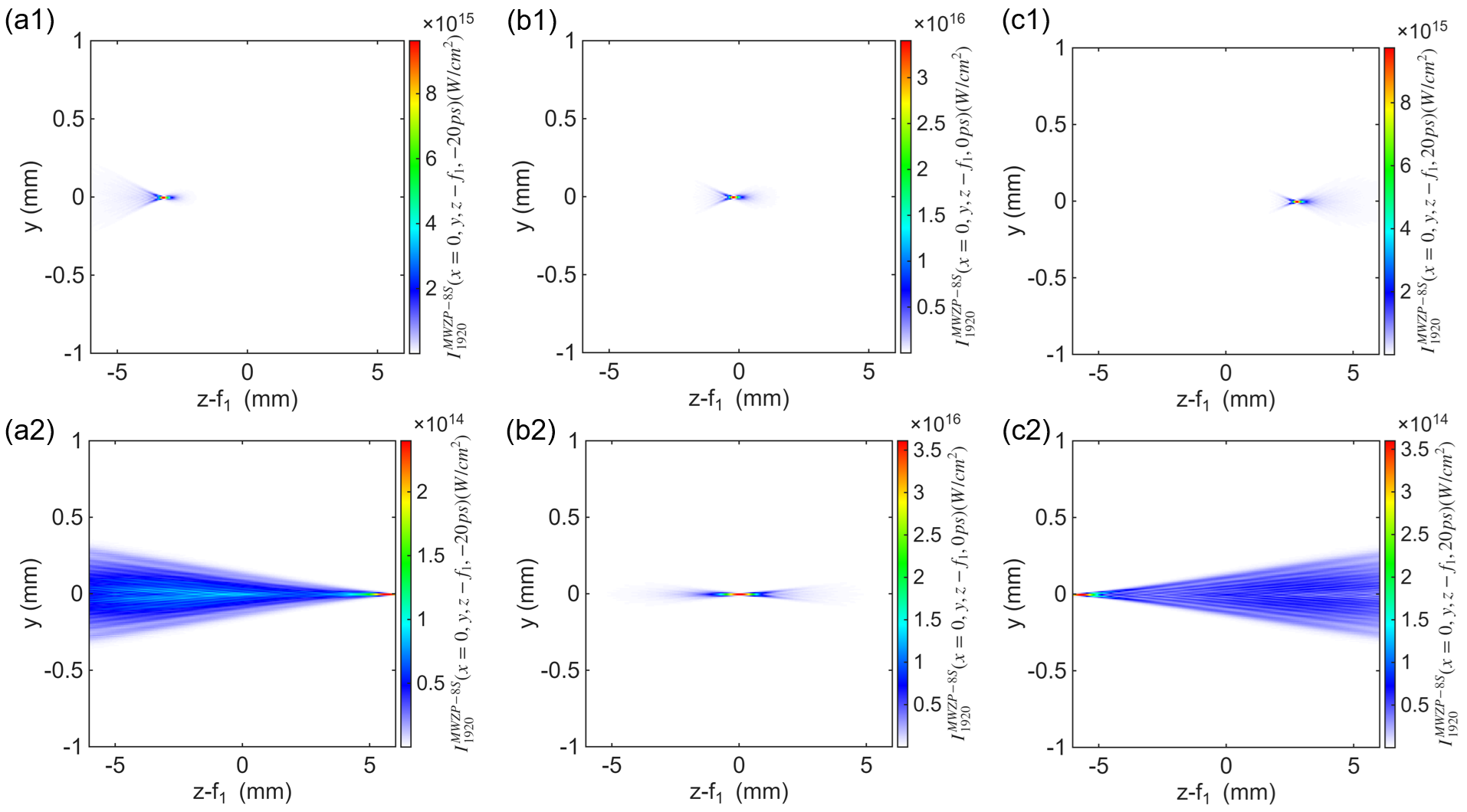}
\caption{Simulated evolution of the flying focus intensity $\mathrm{I}_{1920}^{\mathrm{MWZP-8S}}(\mathrm{x}=0,\mathrm{y},\mathrm{z}-\mathrm{f}_1,\mathrm{t}-\frac{\mathrm{f}_1}{\mathrm{c}})$ for a 115 fs chirped Gaussian pulse with (a1-c1) positive and (a2-c2) negative chirp. Time zero corresponds to the arrival of the 800 nm central wavelength at the focus in the negative chirped case. The positively chirped pulse generates a focus that moves forward at a velocity of 0.5c, traversing the focus range from z $\thicksim$ -3.2 mm to z $\thicksim$ 2.8 mm between -20 ps and 20 ps. Conversely, the negatively chirped pulse produces a focus that propagates backward at a velocity of -1.0c, sweeping from z $\thicksim$ 6 mm to z $\thicksim$ -6 mm between -20 ps and 20 ps.
%{\fcb y-label: $I_{1920}^{MWZP}(x=0,y,z-f_1,t-\frac{f_1}{c})(W/cm^2) $}
}
\label{positive and negative chirp}
\end{figure}

Figure \ref{positive and negative chirp}a1-c1 present the simulated flying-focus dynamics for a positively chirped Gaussian pulse.
In Figure \ref{positive and negative chirp}a1, at $\mathrm{t} = -20$ ps, the chirped pulse is focused at $\mathrm{z} = -3.2$ mm.
At $\mathrm{t} = 0$ ps, the focal position shifts to $\mathrm{z} = -0.2$ mm, as illustrated in Figure \ref{positive and negative chirp}b1.
Subsequently, at $\mathrm{t} = 20$ ps, the focus reaches $\mathrm{z} = 2.8$ mm (Figure \ref{positive and negative chirp}c1).
The extracted focal-spot velocity is $0.5c$, consistent with the value derived from Figure \ref{fig:chirp_velocity}.
Throughout this evolution, the transverse focal-spot size $\sigma_\perp$ remains nearly unchanged.
% Figure \ref{positive and negative chirp}a-c shows the simulation results of the flying focus dynamics for a positively chirped Gaussian pulse. 
% In Figure \ref{positive and negative chirp}a,  
%  where  $\mathrm{t} = -10$ ps ,
% the chirped pulse is  focused at $\mathrm{z} = -3$ mm;
% %{\lzk in the front of the primary focus} $(z\thicksim-3mm)$. 
%  $\mathrm{t} = 0$ ps, focused at $\mathrm{z} = -0.08$ mm (Figure \ref{positive and negative chirp}b);
% and then $\mathrm{t} = 10$ ps, focused at $\mathrm{z} = 2.83$ mm (Figure \ref{positive and negative chirp}c).
% % And then keep moving, at t=??, the focal spot is at z=??? cm  (\ref{positive and negative chirp}c)
% The speed of the focal spot is then calculated to be  0.97c,  which is the same as the velocity obtained in Figure \ref{chirp velocity}.
% During the process, the focal spot size in the transverse direction, $\sigma_\perp$, maintains constant.

\subsubsection{Flying focus with a negative chirp}

Figure \ref{positive and negative chirp}a2-c2 depict the flying-focus dynamics for a negatively chirped Gaussian pulse.
The first snapshot (Figure \ref{positive and negative chirp}a2) shows the laser focusing at $\mathrm{z}\simeq6$ mm, indicating that the shortest-wavelength spectral component of the chirped Gaussian pulse comes to focus first.
The focal spot then travels backward to $\mathrm{z}\simeq0$ mm (Figure \ref{positive and negative chirp}b2), corresponding to longer-wavelength components of the beam.
Afterwards, the longest-wavelength component forms a focus at $\mathrm{z}\simeq -6$ mm, as presented in Figure \ref{positive and negative chirp}c2.
The full axial focal range spans $\Delta \mathrm{z}\simeq12$ mm over a time interval of approximately $\Delta \mathrm{t}\simeq40$ ps, yielding a negative focal velocity of $-1.0c$.
Similar to the positive-chirp scenario, the transverse focal-spot size $\sigma_\perp$ stays constant during propagation for the negative-chirp configuration.
% {\fcb 和前端叙述风格保持基本一致。 
%  Figure \ref{positive and negative chirp}d-f illustrates the flying focus dynamics for a negatively chirped Gaussian pulse. 
%  The first frame (\ref{positive and negative chirp}d)  shows that the laser reaches a focus at  $\mathrm{z}\simeq6$ mm, which means that the shortest-wavelength component of the chirped Gaussian pulse focuses first. 
% Then, the focus moves backward to   $\mathrm{z}\simeq0$ mm (Figure \ref{positive and negative chirp}e), which is a longer-wavelength component of the beam.
% Subsequently, the longest wavelength is focused at $\mathrm{z}\simeq -6$ mm, as illustrated in Figure \ref{positive and negative chirp}f. 
% The total focal length is about  $\Delta \mathrm{z}\simeq12$ mm, which takes time of about  $\Delta \mathrm{t}\simeq40$ ps, corresponding to a negative velocity of -1.0c. 
% The same as the positive chirp case, the focal spot size in the transverse direction of the negative chirp case, $\sigma_\perp$, maintains constant in the propagation process.

\subsubsection{Corresponding space-time ($t-z$) trajectories for different chirps}
\begin{figure}[htbp]
\centering \includegraphics[width=1\linewidth]{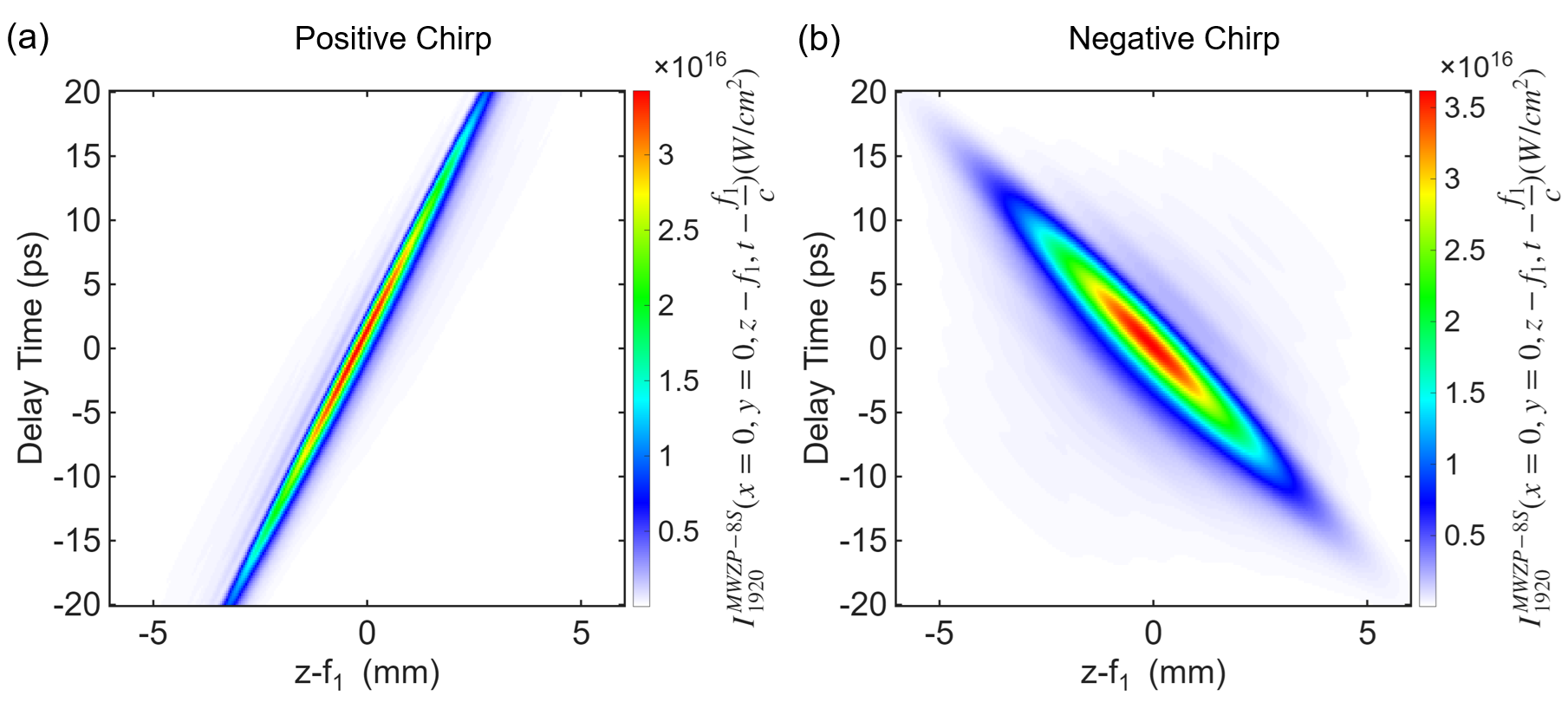}
\caption{
The beam intensity profiles, $\mathrm{I}_{1920}^{\mathrm{MWZP-4S}}(\mathrm{x}=0,\mathrm{y}=0,\mathrm{z}-\mathrm{f}_1,\mathrm{t}-\frac{\mathrm{f}_1}{\mathrm{c}})$,   of the optimized MWZP-4S, showing their dependence on both axial position  $z$ and time $t$. 
%Engineered spatiotemporal  of a flying focus. 
The   Gaussian pulse which has width of 115 fs, wavelength of 800 nm,  and   positive  (a)  or   negative chirp (b) are shown.
The  flying focus speed of $+0.5\mathrm{c}$ (a) and $-1.0\mathrm{c}$ (b) can be obtained.
}
\label{Discussion}
\end{figure}

% {\lzk 讨论会议新发现的公式The flying focus scheme has two  important characteristics: the decoupling of focal velocity from the group velocity and extended focal range, 
% which can be recognised in Figure \ref{Discussion}, where the 2D plot of  the intensity $\mathrm{I}_{1920}^{\mathrm{4MWZP}}(\mathrm{x}=0,\mathrm{y}=0,\mathrm{z}-\mathrm{f}_1,\mathrm{t}-\frac{\mathrm{f}_1}{\mathrm{c}}) $ is shown. 

% In Figure \ref{Discussion}(a), 
% with a positive chirp, the focus moves forward at a velocity of $+0.97\mathrm{c}$, 
% sweeping the range of   $-6<\mathrm{z}< 6$ mm. 
% Conversely, in Figure \ref{Discussion}(b), with a negative chirp, the focus flies backward at -1.0c, covering the same range, $-6<\mathrm{z}<6$ mm.
% Both of them exhibit the decoupling of the focal velocity from the group velocity. 
% By appropriately arranging the chirp of the incident light pulses, the focal spot can propagate at any velocity.
% The extended focal region is about 12 mm, exceeding 140 Rayleigh ranges. }
The independence of focal velocity with respect to group velocity and an extended focal range are crucial parameters for the flying focus. Figure \ref{Discussion} visualizes these parameters via the two-dimensional intensity distribution $\mathrm{I}_{1920}^{\mathrm{MWZP-8S}}(\mathrm{x}=0,\mathrm{y},\mathrm{z}-\mathrm{f}_1,\mathrm{t}-\frac{\mathrm{f}_1}{\mathrm{c}})$.

For the positive-chirp case in Figure \ref{Discussion}(a), the focal spot advances at $+0.5\mathrm{c}$ throughout the axial range of $-3.2<\mathrm{z}<2.8$ mm. For negative chirp in Figure \ref{Discussion}(b), the focal spot recedes at $-1.0\mathrm{c}$ across the range of $-6<\mathrm{z}<6$ mm.
Both cases confirm the effective decoupling between focal propagation velocity and pulse group velocity. Flexible tuning of the incident chirp enables arbitrary manipulation of focal moving speed. The resulting extended focal region covers approximately 12 mm, exceeding 140 Rayleigh lengths.

Figures \ref{Discussion}(a) and (b) illustrate that the velocity deviation remains small under positive chirp, whereas a much larger deviation emerges for negative chirp. Statistics over all slopes can be interpreted as a projection onto the propagation direction, and the width of this projection characterizes the velocity distribution.

% All points lying on a straight line with identical slope correspond to the same flying-focus velocity, while the averaged velocity within each subplot stays constant.

The formula\cite{Spatial_and_temporal_2019}
\begin{equation}
\Delta t = \frac{2Z_r}{v_g}\left| \frac{v_g}{v_f} - 1 \right| + \frac{R}{2v_g}\left(\sqrt{(f^\#)^2 + 1/4} - f^\#\right)
\label{eq:ontimewidth}
\end{equation}
gives the on-axis temporal width, namely the effective pulse duration within the longitudinal focal region.
Here, $Z_r = 4\lambda_0(f^\#)^2/\pi$ denotes the Rayleigh length, and the $f$-number is defined as $f^\# = f_0/(2R)$.
% The mathematical explanation of the situation attributes to the Eq.\ref{eq:ontimewidth}.
The second term of Eq.\ref{eq:ontimewidth} describes the wavefront tilt contribution from the geometric path length difference between the central ray and the most radial ray. For Gaussian pulses with different chirps passing through the same optimized zone plate, this wavefront-tilt term exhibits negligible variation. Accordingly, the difference in on-axis temporal width predominantly originates from the first term. 
This substantial discrepancy in the magnitude of $\left|v_g/v_f -1\right|$ yields a marked difference in the on-axis temporal width. For the positive-chirp flying focus with $v_f=0.5c$, this factor equals $1$, while the negative-chirp case with $v_f=-1.0\mathrm{c}$ gives a value of $2$. Consequently, the positive-chirp configuration exhibits a much narrower on-axis temporal width, whereas the negative-chirp counterpart produces a considerably longer on-axis pulse duration.
% This leads to a narrower on-axis temporal width for the positive-chirp case and a longer temporal width for the negative-chirp configuration.

% {\blue Beyond these scheme-level characteristics, the MWZP design itself, as a binary optical element, offers distinct practical advantages: }
% it is highly compatible with modern nanofabrication, which requires only a single lithography step followed by metal deposition and lift-off, thereby avoiding complex processes, such as etching and multilayer alignments. Therefore, in the experiment, when the required diffractive lense's radius is too large to be obtained in the market or the lense is destroyed by a strong laser, the binary MWZP is an ideal alternative as more zone pairs can be easily added during the fabrication process.

% The applications of Phase MWZP or Four-zone Phase MWZP, with enhanced focusing efficiency, can perform in more scenes that require stronger fluences compared with binary MWZP.成本相对低廉 可以作为飞行焦点的预实验
% 该波带片结合尾波加速(LWFA)的电子可以用于低成本产生EUV光源的预实验方案。\cite{gerstmayr2026experimentaldemonstrationflyingfocusenhanced}

\section{Discussion}

\begin{figure}[htbp]
\centering 
\includegraphics[width=1\linewidth]{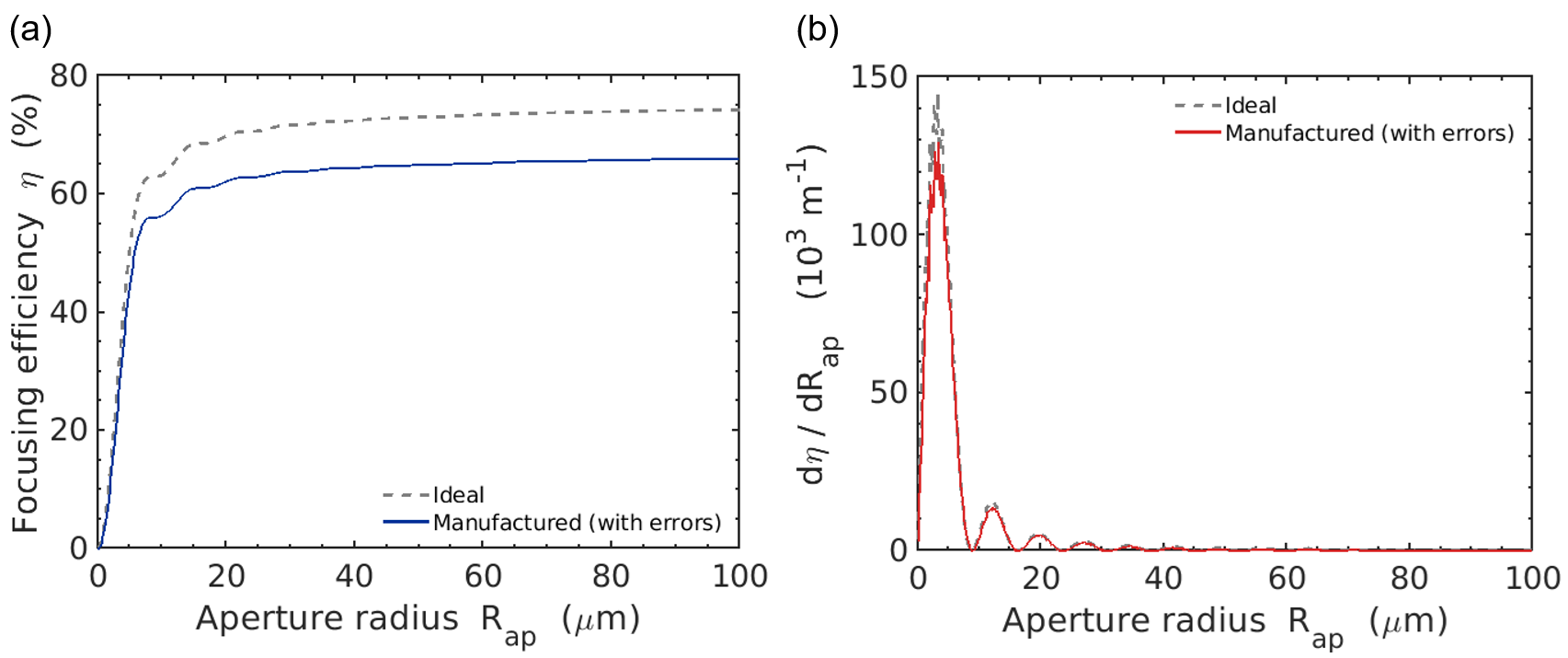}
\caption{(a) The focusing efficiency of the 1920-zone, MWZP-8S as a function of the aperture radius. The dashed grey line represents the ideal theoretical design, while the solid blue line depicts the performance of the manufactured plate with a random etching depth error of \(\pm 5\%\). (b) The derivative of the focusing efficiency, \(d\eta/dR_{ap}\), as a function of the aperture radius for the ideal and manufactured cases. The peaks correspond to the maximum energy accumulation rates of the primary focal spot.}
\label{fabrication_stability}
\end{figure}

For a transmissive phase plate, the phase delay \(\phi(r)\) introduced by the local etch depth \(d(r)\) is dictated by the refractive index contrast between the substrate and the surrounding medium. Specifically, for our operating wavelength of \(\lambda = 800\ \mathrm{nm}\), using fused silica (\(n_{\mathrm{mat}} = 1.4533\)\cite{Malitson1965}) in air (\(n_{\mathrm{env}} = 1.0000\)), the phase transmittance \(p(r) = \phi(r)/\pi\) is directly proportional to the etch depth:
\begin{equation}
d(r) = \frac{\lambda}{2\left( n_{\mathrm{mat}} - n_{\mathrm{env}} \right)} \, p(r) = d_{\pi} \, p(r).
\label{eq:etch-depth-discussion}
\end{equation}
For the MWZP-8S, the phase steps are refined to \(\pi/4\) increments, meaning the corresponding etch depth difference between adjacent steps is \(d_{\pi}/4 \approx 220.6\ \mathrm{nm}\). Such small depth quantization imposes stringent requirements on the fabrication process.

In practical multi-step lithography and etching processes, fluctuations in etching rates and material uniformity inevitably introduce random etching depth deviations. To evaluate the robustness of our design against fabrication imperfections, we simulate the manufactured plate by introducing a random \(\pm 5\%\) etching depth error to the ideal 8-step profile. As illustrated in Fig.~\ref{fabrication_stability}(a), the focusing efficiency of the manufactured plate closely follows the ideal curve. However, due to the phase mismatch induced by the depth error, the saturation efficiency decreases from the ideal 74.3\% to 66.1\%, with a drop of 11\%. 

This degradation is also revealed in the derivative curves shown in Fig.~\ref{fabrication_stability}(b). The $\pm 5\%$ depth error distorts the ideal phase-matching condition, causing energy leakage from the primary focus. Consequently, the peak energy accumulation rate also drops by approximately \(11\%\).
% Consequently, the peak energy accumulation rate drops from $145.0 \times 10^{3}~\mathrm{m}^{-1}$ to $129.2 \times 10^{3}~\mathrm{m}^{-1}$. 
Nevertheless, the dominant focal spot remains highly concentrated, and the energy is still overwhelmingly confined within the first few micrometers. These results demonstrate that the MWZP-8S possesses excellent robustness against realistic etching depth errors, maintaining a strong focusing performance and ensuring its practical feasibility in high-intensity laser applications.

 % Although the standard multilevel FZPs can have higher efficiency more than 90\%, it is not selected for further comparison in this study. As derived in Eq. (3) of \cite{DiFabrizio1999}, an \(n\)-level phase plate is fundamentally unable to suppress the \((n+1)\)th-order focus. Because the summation term \(\sum \exp[i2\pi\frac{l-1}{L}(m-1)]\) only vanishes when the diffraction order \(m\) does not match the number of levels \(L\), the residual energy from unwanted higher-order foci inevitably introduces unwanted background noise, which severely degrades the contrast of the primary focus.

The MWZP-$2^{M}$S, owing to their enhanced focusing efficiency, can be applied in a broader range of scenarios requiring higher fluence than the binary MWZP. Their relatively low cost, with the help of modern nanofabrication, also makes them suitable for preliminary experiments on flying focus. When combined with electrons from laser wakefield acceleration (LWFA), such zone plates can enable a low-cost proof-of-principle scheme for EUV light source generation\cite{gerstmayr2026experimentaldemonstrationflyingfocusenhanced}.
 % At X-ray wavelengths where standard optical refractive lenses fail本身设计思路就是从X射线波段来的，自然可以考虑生成X波段的飞行焦点\cite{DiFabrizio1999}, zone plates allow precise wavefront manipulation\cite{Morillo-Candas_2026}, showing strong potential for implementing flying focus in the X-ray regime.
 Because the design rationale itself originates from the X-ray band\cite{DiFabrizio1999}, it is also natural to consider producing a flying focus in the X-ray regime. At such wavelengths, where standard refractive lenses become ineffective, zone plates enable precise wavefront control\cite{Morillo-Candas_2026}, which makes them highly promising for realizing flying-focus schemes in the X-ray region.
% Ps-level pulses enabling superluminal ionization fronts can greatly boost terahertz energy\cite{PhysRevLett.134.045001}, beneficial for remote spectroscopy\cite{Talbi_2023} and terahertz imaging\cite{Zhao_2014}. 

% {\lzk To evaluate the influence of phase quantization, we also examined an alternative four-zone phase MWZP design. In this design, the corresponding complex-amplitude transmittances are $t_{1}=\exp(i\pi/2)$, $t_{2}=\exp(i2\pi)$, $t_{3}=\exp(i3\pi/2)$, and $t_{4}=\exp(i\pi)$. Numerical verification gives a focusing efficiency of $47.8\%$ for $N=1000$. We further varied the phase assignments of the four zones and found that the focusing efficiency remains comparable as long as the radial phase profile preserves a stepwise decreasing, grating-like distribution. Among these configurations, the phase profile used in Fig.~4 provides the best performance, with a focusing efficiency of $49.5\%$ for $N=1000$; therefore, this configuration was selected for the present work.
% }

% ========================================
\section{Summary}

In this study, by modulating the zone phase in multiple discrete step, the number of zone plates, and the chirp parameters of incident Gaussian pulses, 
various multilevel MWZPs have been designed for  flying focus applications, which are characterized by enhanced focusing efficiency, deeply suppressed higher-order foci, and a controllable focal speed.
% highly extended focal range.
% a modulated width zone plate (MWZP) with easy fabrication via modern nanofabrication technology and cost-effectiveness,as well as deeply suppressed high-order foci and controllable focal speed,has been designed for flying focus applications.

The MWZP-2S replaces opaque zones with \(\pi\) phase-shifting zones; constructive interference yields a fourfold focal peak intensity enhancement over the MWZP. With finer four-step phase segmentation, the MWZP-4S achieves a focal peak intensity \(9.5\times\) that of the MWZP. Most notably, the MWZP-8S pushes the focal peak intensity to \(12\times\) that of the MWZP and achieves a saturated aperture efficiency of \(73.7\%\). Numerical simulations confirm that 
the intensities of higher-order foci up to the 25th order are suppressed by four orders of magnitude with a MWZP-8S of 1000 zones.
This single-focus characteristic becomes pronounced when sufficient annular zones are illuminated, which suppresses high noises far from the primary focus. 
With an appropriate chirp parameter for the incident Gaussian pulse, the flying focus can propagate at any forward or backward velocity-whether smaller than the speed of light or exceeding it-over an extended focal region of more than 100 Rayleigh ranges.
% With an appropriate chirp parameter for the incident Gaussian pulse, the flying focus can propagate with any forward or backward velocity, whatever smaller than speed of light or larger than the speed of light,  with an extended focal region over 100 Rayleigh ranges.

The proposed phase MWZP method may offer significant potential across multiple areas, including 
enhancing  EUV radiation sources\cite{gerstmayr2026experimentaldemonstrationflyingfocusenhanced} and paving the way for implementing flying focus concepts in spectral regions such as X-rays\cite{DiFabrizio1999}.
 
 % high-efficiency laser amplifiers\cite{PhysRevLett.120.024801}, 
 % TeV laser wakefield accelerators\cite{10.1063/5.0274780,PhysRevLett.124.134802}, 
 % monochromatic EUV sources\cite{ComptonZhao2025},
 % and more.

% --- 正文结束，附录开始 ---
\appendix % 关键命令

\section{Fabrication consideration of the MWZP-4S} 
\begin{figure}
    \centering
    \includegraphics[width=1\linewidth]{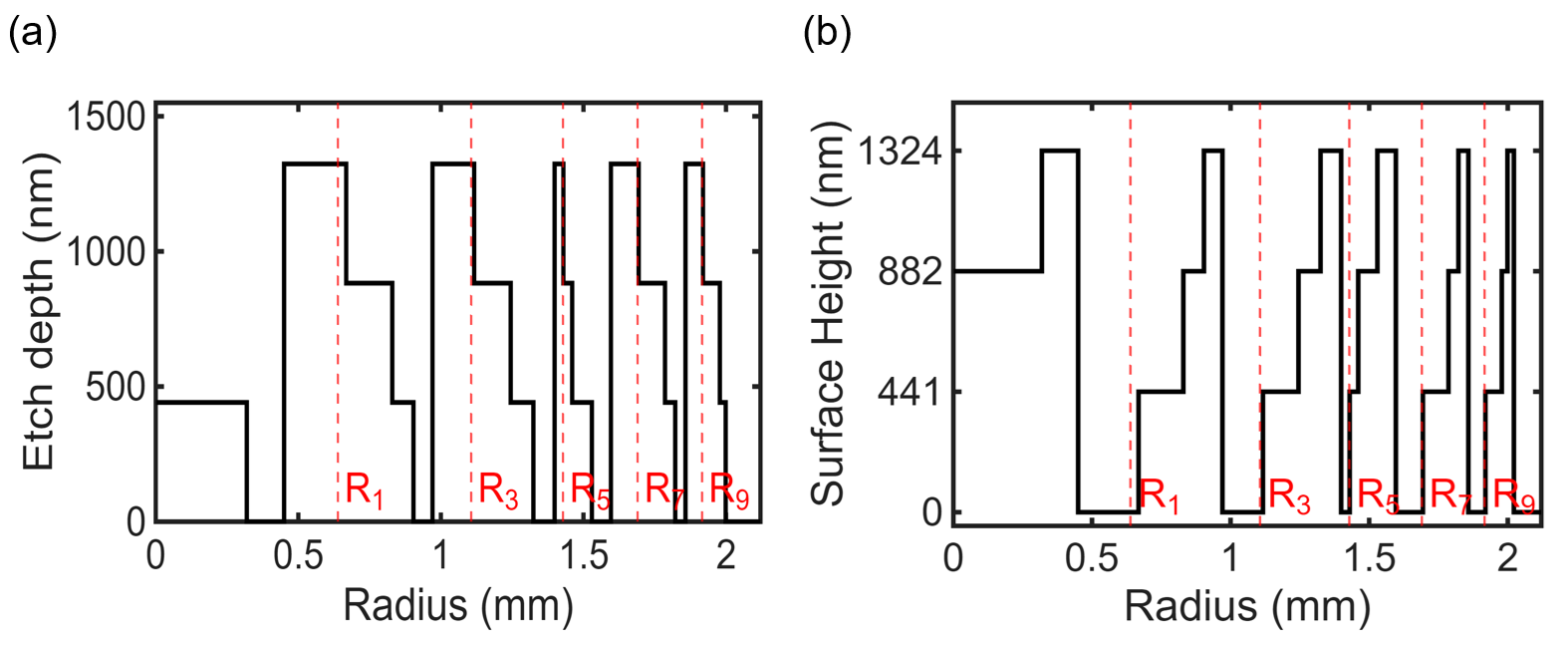}
    \caption{The fabrication consideration of MWZP-4S using ultraviolet photolithography and overlay lithography. (a) The graph depicting the radial etch depth  of the MWZP-4S as a function of its radius.(b) The surface profile  of the MWZP-4S as a function of its radius. }
    \label{4zone phase mwzp depth}
\end{figure}
\subsection{Radial etch depth calculation}
\label{app:etch-depth}

For a transmissive phase plate, the phase delay $\phi(r)$ introduced by the local etch depth $d(r)$ is given by
\begin{equation}
\phi(r) = \frac{2\pi}{\lambda} \left( n_{\mathrm{mat}} - n_{\mathrm{env}} \right) d(r),
\label{eq:phase-depth}
\end{equation}
where $\lambda$ is the operating wavelength, $n_{\mathrm{mat}}$ is the refractive index of the substrate material, and $n_{\mathrm{env}}$ is the refractive index of the surrounding medium. In our design, the operating wavelength is $\lambda = 800\ \mathrm{nm}$, the substrate is fused silica with $n_{\mathrm{mat}} = 1.4533$ \cite{Malitson1965}, and the surrounding medium is air with $n_{\mathrm{env}} = 1.0000$.

In the phase profile, the phase transmittance is expressed in units of $\pi$, i.e., $p(r) = \phi(r)/\pi$. The four phase levels are
\begin{equation}
p(r) = 0,\ 0.5,\ 1,\ 1.5,
\end{equation}
corresponding to $0,\ \pi/2,\ \pi,\ 3\pi/2$, respectively. From Eq.~\eqref{eq:phase-depth}, the etch depth can be written as
\begin{equation}
d(r) = \frac{\lambda}{2\left( n_{\mathrm{mat}} - n_{\mathrm{env}} \right)} \, p(r) = d_{\pi} \, p(r).
\label{eq:etch-depth}
\end{equation}

Thus, the four phase levels correspond to the etch depths
\begin{equation}
d(r) = 0,\ 441.21,\ 882.42,\ 1323.63\ \mathrm{nm},
\label{eq:four-depths}
\end{equation}
for $p(r) = 0,\ 0.5,\ 1,\ 1.5$, respectively. The maximum etch depth is $d_{\mathrm{max}} = 1323.63\ \mathrm{nm}$, which corresponds to the $3\pi/2$ phase level. 

The four-level structure is implemented by two binary overlay lithography steps. The first etching step has a depth of approximately $441.21\,\mathrm{nm}$, and the second overlay etching step has a depth of approximately $882.42\,\mathrm{nm}$. Depending on the combination of the two binary masks, the superposition of these two steps yields four depth levels: $0$, $441.21$, $882.42$, and $1323.63\,\mathrm{nm}$.

\subsection{Surface relative height after UV lithography and overlay lithography}
\label{app:surface-height}

After ultraviolet (UV) lithography and overlay lithography, the final surface profile is determined by the local etch depth. To characterize the fabricated surface, we define the relative surface height $h_{\mathrm{rel}}(r)$ with respect to the deepest etched groove, i.e., the bottom of the $3\pi/2$ phase level at $d_{\mathrm{max}} = 1323.63\ \mathrm{nm}$. The relative height is defined as
\begin{equation}
h_{\mathrm{rel}}(r) = d_{\mathrm{max}} - d(r).
\label{eq:rel-height}
\end{equation}
Substituting Eq.~\eqref{eq:etch-depth} into Eq.~\eqref{eq:rel-height}, we obtain
\begin{equation}
h_{\mathrm{rel}}(r) = d_{\mathrm{max}} - d_{\pi} \, p(r)
                    = 1323.63 - 882.42 \, p(r) \quad (\mathrm{nm}).
\label{eq:rel-height-explicit}
\end{equation}
Therefore, the four phase levels correspond to the relative surface heights
\begin{equation}
h_{\mathrm{rel}}(r) = 1323.63,\ 882.42,\ 441.21,\ 0\ \mathrm{nm},
\label{eq:four-heights}
\end{equation}
for $p(r) = 0,\ 0.5,\ 1,\ 1.5$, respectively. The deepest etched region ($p=1.5$) has $h_{\mathrm{rel}} = 0\ \mathrm{nm}$, while the unetched surface ($p=0$) has $h_{\mathrm{rel}} = 1323.63\ \mathrm{nm}$. These values are consistent with the radial etch depth profile in Fig.~\ref{4zone phase mwzp depth}(a) and the surface relative height profile in Fig.~\ref{4zone phase mwzp depth}(b).
Meanwhile, the surface relative height profile is consistent with the quaternary Fresnel zone plate\cite{DiFabrizio1999}.

\section{The transfer function of the transverse components of the envelope}

The transverse components' equation of the chirped Gaussian pulse
$\mathrm{E}_{\perp}$\cite{PhysRevA.97.033835},
\begin{equation}
    E_{\perp}(\vec{r},\xi,z)=-\frac{ik_{0}-\partial_{\xi}}{2\pi z}\int \exp\left[\frac{ik_{0}}{2z}(\vec{r} - \vec{r}')^2\right] \cdot E_{\perp}\left(\vec{r'},\xi-\frac{\left| \vec{r} - \vec{r}' \right|^2}{2z},0\right)dr',
     \label{transverse electric fields2}
\end{equation}
can be understood as describing the field in the transverse plane at any z, which results from the superposition of light emitted from a collection of point sources in the plane $z=0$.
Here $\vec{\mathrm{r}}$ signifies the location for calculating the diffraction intensity, $\vec{\mathrm{r}'}$ represents the incident zone plate whose intensity is already known, and z is the propagation distance on the optical axis. $\xi=\mathrm{ct-z}$ is the moving frame coordinate.
$\mathrm{E}_{\perp}(\mathrm{r}',\xi,0)$ determine the electric field's amplitudes and phases at $\mathrm{z}=0$.
The term $\frac{|\vec{\mathrm{r}}-\vec{\mathrm{r}'}|^2}{2\mathrm{z}}$ translating $\xi$ term considers the time differences of light propagating from different points on the designed zone plate. 

The precise analytical solution for Eq. \ref{transverse electric fields2} at the propagation axis, specifically addressing the spatiotemporal Gaussian pulse profile, is
\begin{equation}
E_1(\mathbf{r}, \xi, 0) = E_a \exp\left[
-\frac{r^2}{w_0^2} 
- \frac{\xi^2}{c^2 \tau^2 (1 - i\eta)}
\right].
\label{electric fields0}
\end{equation}
A more comprehensible form of $\mathrm{E}_{\perp}$ is obtained by plugging Eq. \ref{electric fields0} into Eq. \ref{transverse electric fields2},
\begin{equation}
\tilde{E}_{\perp} \left( \vec{r}, \xi, z \right) = -\frac{\left( i k_0 - \partial_{\xi} \right)}{2 \pi z} \int \exp\left[ \frac{i k_0}{2 z} \left| \vec{r} - \vec{r}' \right|^2 \right] \cdot E_a \exp\left[ -\frac{r'^2}{w_0^2} - \frac{\left( \xi - \frac{\left| \vec{r} - \vec{r}' \right|^2}{2 z} \right)^2}{c^2 \tau^2 (1 - i \eta)} \right] \mathrm{d}\vec{r}'.
\label{electric fields_simplify1}
\end{equation}
The assistant parameter, which contains the chirp term, is defined as,
\begin{equation}
A = c^2 \tau^2 (1 - j \cdot \eta).
\end{equation}
Then Eq. \ref{electric fields_simplify1} can be simplified to be
\begin{equation}
\tilde{E}_{\perp} \left( \vec{r}, \xi, z \right) = -\frac{\left( i k_0 - \partial_{\xi} \right)}{2 \pi z} \int \exp\left[ \frac{i k_0}{2 z} \left| \vec{r} - \vec{r}' \right|^2 \right] \\
\cdot E_a \exp\left[ 
-\frac{r'^2}{w_0^2} 
- \frac{1}{A} \left( 
  \xi^2 + \frac{|\vec{r} - \vec{r}'|^4}{4z^2} 
  - \frac{\xi}{z} |\vec{r} - \vec{r}'|^2
\right)
\right].
\label{electric fields_simplify2}
\end{equation}

 The transverse  electric fields of  the chirped Gaussian pulse are then split as a convolution of the Gaussian light beams with impulse response functions. 
Here is the complete transfer function in the modified Fresnel integrals,
\begin{equation}
h = \frac{1}{2 \pi z} \left( -j k_0 - \frac{2 \xi}{A} + \frac{\left| \vec{r} - \vec{r}' \right|^2}{Az } \right) \cdot \exp\left( \frac{j k_0}{2 z}\left| \vec{r} - \vec{r}' \right|^2 + \frac{\xi \left| \vec{r} - \vec{r}' \right|^2}{A z} - \frac{\left| \vec{r} - \vec{r}' \right|^4}{4 A z^2} \right).
\end{equation}

The beam part of the convolution is then determined to be,
\begin{equation}
U_1 = t_{plate} \cdot E_0 \cdot \exp\left(-\frac{r^2}{w_0^2} - \frac{\xi^2}{A}\right),
\label{input_beam}
\end{equation}
which is a Gaussian wave with $\xi$ term that passes through the zone plate.

\section{Mathematical derivation of the local relative distance approximation used in section 2.2.A }
\label{app:delta_derivation}

This appendix provides the detailed mathematical derivation of the approximation for the local relative distance \(\delta\), which bridges the localized linear distance and the global zone phase within the \(n\)-th zone pair.

By the definition of a Fresnel zone plate, the radius of the \(m\)-th zone is given by:
\begin{equation}
R_m = \sqrt{m\lambda f}.
\label{Rm}
\end{equation}
The square of the first zone radius is \(R_1^2 = \lambda f\). Therefore, the squared radius of the \((2n-1)\)-th zone, which corresponds to the center of the \(n\)-th zone pair, can be expressed as:
\begin{equation}
R_{2n-1}^2 = (2n-1)\lambda f = (2n-1)R_1^2.
\end{equation}

\textbf{Step 1: Approximation of the denominator (half-width of the zone pair).} \\
For a sufficiently large \(n\), the radial difference between adjacent zones becomes infinitesimally small. Thus, the denominator \(R_{2n} - R_{2n-1}\) can be approximated using the derivative with respect to the zone index \(m\):
\begin{equation}
R_{2n} - R_{2n-1} \approx \left. \frac{dR}{dm} \right|_{m=2n-1} = \frac{d}{dm}\left(\sqrt{m\lambda f}\right) = \frac{\lambda f}{2\sqrt{m\lambda f}} = \frac{R_1^2}{2R_{2n-1}}.
\end{equation}

\textbf{Step 2: Approximation of the numerator (local radial distance).} \\
Within the \(n\)-th zone pair, the radial coordinate \(r\) is very close to \(R_{2n-1}\). Utilizing the difference of squares, we have:
\begin{equation}
r^2 - R_{2n-1}^2 = (r - R_{2n-1})(r + R_{2n-1}).
\end{equation}
Since \(r \approx R_{2n-1}\), it follows that \(r + R_{2n-1} \approx 2R_{2n-1}\). The numerator can therefore be approximated as:
\begin{equation}
|r - R_{2n-1}| = \frac{|r^2 - R_{2n-1}^2|}{r + R_{2n-1}} \approx \frac{|r^2 - R_{2n-1}^2|}{2R_{2n-1}}.
\end{equation}

\textbf{Step 3: Assembly of the relative distance \(\delta\).} \\
Substituting the results from Steps 1 and 2 into the definition of \(\delta = |r - R_{2n-1}| / (R_{2n} - R_{2n-1})\), we obtain:
\begin{equation}
\delta \approx \frac{\frac{|r^2 - R_{2n-1}^2|}{2R_{2n-1}}}{\frac{R_1^2}{2R_{2n-1}}} = \frac{|r^2 - R_{2n-1}^2|}{R_1^2}.
\end{equation}
By expanding \(R_{2n-1}^2\), the approximation yields:
\begin{equation}
\delta \approx \left| \frac{r^2}{R_1^2} - (2n-1) \right|.
\end{equation}

Within the \(n\)-th zone pair, \(r\) ranges from \(R_{2n-2}\) to \(R_{2n}\), which implies that \(r^2/R_1^2\) spans from \(2n-2\) to \(2n\). This indicates that the sign of the expression \(\frac{r^2}{R_1^2} - (2n-1)\) alternates (i.e., negative before the zone pair center and positive after). This property is essential for the trigonometric reduction that equates the expected transmittance of the MWZP to the continuous transmittance of a negative GZP in Eq.\ref{approxGZP}.

 \section{Mathematical derivation of the 1.028 factor for evaluating focusing efficiency}
\label{appendix: fwhm_derivation}

To precisely evaluate the focusing capability of the proposed zone plates, it is essential to establish a rigorous theoretical reference for the FWHM of the focal spot. According to the Fresnel-Kirchhoff diffraction theory, the intensity distribution of the focal spot for a circular aperture (or an ideal Fresnel zone plate) in the focal plane follows the Airy pattern, which can be expressed as
\begin{equation}
I(\theta) = I_0 \left[ \frac{2 J_1(k a \sin\theta)}{k a \sin\theta} \right]^2,
\label{eq:airy_intensity}
\end{equation}
where \(I_0\) is the peak intensity at the center, \(J_1(\cdot)\) is the first-order Bessel function of the first kind, \(k = 2\pi/\lambda\) is the wavenumber, \(a\) is the radius of the aperture, and \(\theta\) is the diffraction angle. 

To find the FWHM, we define a dimensionless variable \(u = k a \sin\theta\). The normalized intensity profile is then given by \(I(u)/I_0 = \left[ 2J_1(u)/u \right]^2\). By solving the equation \(I(u) = 0.5 I_0\) numerically, we obtain the first positive root
\begin{equation}
u_{1/2} \approx 1.61633.
\label{eq:half_max_root}
\end{equation}
Under the paraxial approximation (\(\sin\theta \approx \theta\)), the half-width at half maximum (HWHM) in the angular domain is \(\theta_{1/2} = u_{1/2} \lambda / (2\pi a)\). Consequently, the full width at half maximum (FWHM) in the focal plane at a focal length \(f\) is
\begin{equation}
\text{FWHM} = 2 f \theta_{1/2} = \frac{2 \times 1.61633 \lambda f}{2\pi a} \approx 1.028 \frac{\lambda f}{2a}.
\label{eq:fwhm_angular}
\end{equation}
Meanwhile, the Rayleigh criterion (radius of the first dark ring) corresponds to the first zero of \(J_1(u)\), which is \(u_1 \approx 3.8317\). This yields the well-known Rayleigh radius \(r_{\text{Rayleigh}} \approx 1.22 \lambda f / (2a)\)\cite{zhao2008optics}.

For a Fresnel zone plate, the outermost zone width \(\delta_{RN}\) is geometrically related to the aperture radius \(a\) and focal length \(f\) by the relation \(\delta_{RN} \approx \lambda f / (2a)\). By substituting this into Eq. \ref{eq:fwhm_angular}, we obtain the theoretical FWHM reference for an ideal zone plate:
\begin{equation}
\text{FWHM}_{\text{theoretical}} = 1.028 \delta_{RN}.
\label{eq:fwhm_theoretical}
\end{equation}
This derivation mathematically validates the \(1.028\) factor used as the FWHM reference for evaluating focusing efficiency in Section 3.2. 
% The explicit separation between the FWHM criterion (\(1.028\delta_{RN}\)) and the Rayleigh criterion (\(1.22\delta_{RN}\)) ensures that the measured focal spot sizes are rigorously and accurately evaluated against the diffraction limit.

\section{The chirp parameters control the focus velocity}

\begin{figure}[htbp]
\centering
\includegraphics[width=0.6\linewidth]{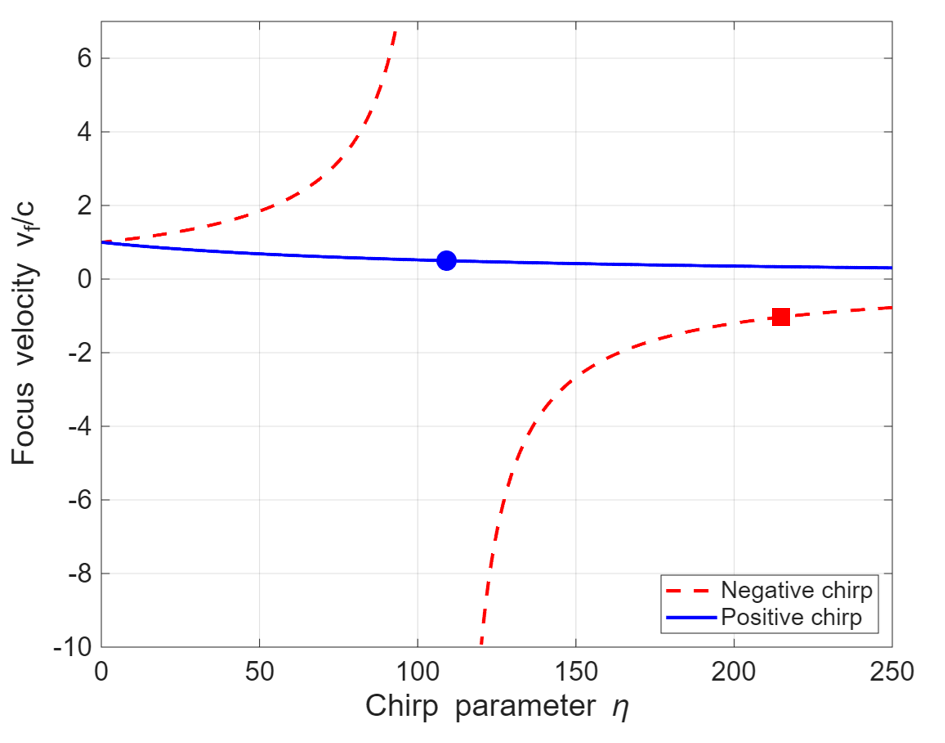}
\caption{The focal velocity of the flying focus pulse as a function of chirp parameter. A positively chirped pulse is limited to positive subluminal focal velocities, while a negatively chirped pulse can have either a positive or negative velocity. The specific cases simulated in the following sections are marked with a blue circle (focal velocity \(+0.5\mathrm{c}\)) and a red square (focal velocity \(-1.0\mathrm{c}\)).}
\label{fig:chirp_velocity}
\end{figure}

\begin{backmatter}
\bmsection{Funding Statement}
National Natural Science Foundation of China (12235003); the National Key R$\&$D Program of China (2023YFA1606900, 2022YFA1602301).

\bmsection{Data Access Statement}
Data underlying the results presented in this paper are not publicly available at this time but may be obtained from the authors upon reasonable request.

\bmsection{Conflict of Interest declaration}
The authors declare that they have no conflicts of interest.
\end{backmatter}

%%%%%%%%%% If using BibTeX:

% \begin{thebibliography}{99}
% \newcommand{\enquote}[1]{``#1''}

% \end{thebibliography}

% \begin{thebibliography}{99}
% \newcommand{\enquote}[1]{``#1''}

\end{CJK*}
\end{document}